\documentclass{article}

\PassOptionsToPackage{numbers, compress}{natbib}

\usepackage[final]{neurips_2022}

\usepackage[utf8]{inputenc} 
\usepackage[T1]{fontenc}    
\usepackage{hyperref}       
\usepackage{url}            
\usepackage{booktabs}       
\usepackage{amsfonts}       
\usepackage{nicefrac}       
\usepackage{microtype}      
\usepackage{xcolor}         

\usepackage{graphicx}
\usepackage{epsfig}
\usepackage{amsmath}
\usepackage{amssymb}

\usepackage{amsmath,amsfonts,bm}

\def\eqref#1{Eq.~\ref{#1}}

\def\1{\bm{1}}

\def\vh{{\bm{h}}}

\def\vy{{\bm{y}}}

\DeclareMathAlphabet{\mathsfit}{\encodingdefault}{\sfdefault}{m}{sl}
\SetMathAlphabet{\mathsfit}{bold}{\encodingdefault}{\sfdefault}{bx}{n}

\usepackage{amsthm}
\usepackage{etoolbox}
\usepackage{siunitx}        
\robustify\bfseries
\usepackage{bm}
\usepackage{color}
\usepackage{extarrows}
\usepackage{subfigure}
\usepackage{multirow}
\usepackage{mathtools}
\usepackage{bbm}
\usepackage{xspace}

\newcommand{\tdash}{\multicolumn{1}{c}{-}}

\newcommand{\vzeta}{\bm{\zeta}}

\newcommand{\fullname}{Audio Separator and Motion Predictor\xspace}
\newcommand{\name}{ASMP\xspace} 

\newcommand{\vid}{V}
\newcommand{\point}{\textbf{x}}
\newcommand{\frmwind}{\ell}
\newcommand{\frm}{F}
\newcommand{\disp}{\mathbf{d}}
\newcommand{\src}{\mathbf{a}} 
\newcommand{\srcspec}{\mathbf{S}}
\newcommand{\nodes}{\mathcal{V}}
\newcommand{\node}{n} 
\newcommand{\edge}{e}
\newcommand{\edges}{\mathcal{E}}
\newcommand{\graph}{\mathcal{G}}

\newcommand{\pobjs}{P}
\newcommand{\pobj}{p}
\newcommand{\objclasses}{\mathcal{C}}

\newcommand{\loss}{\mathcal{L}}

\newcommand{\feat}{F}
\newcommand{\embed}{\vy}
\newcommand{\embedset}{Y}
\newcommand{\mspec}{\bf{X}}
\newcommand{\mask}{\bf{M}}

\newcommand{\real}{\mathbb{R}}
\newcommand{\reals}[1]{\mathbb{R}^{#1}}

\newcommand{\set}[1]{\left\{#1\right\}}

\DeclareMathOperator{\frcnn}{FRCNN}

\newcommand{\etal}{\textit{et al.}}

\usepackage{tikz}
\usepackage{comment}

\usepackage[accsupp]{axessibility}  

\title{Learning Audio-Visual Dynamics Using Scene Graphs \\for Audio Source Separation}

\author{
  Moitreya Chatterjee$^{1*}$ \qquad\qquad Narendra Ahuja$^{1}$ \qquad\qquad Anoop Cherian$^{2}\thanks{Equal Contribution.}$\\
  \texttt{metro.smiles@gmail.com} \qquad \texttt{n-ahuja@illinois.edu} \qquad \texttt{cherian@merl.com }  \\
  $^1$University of Illinois, Urbana-Champaign, Urbana, IL \\ $^2$Mitsubishi Electric Research Labs, Cambridge, MA\\
}

\begin{document}

\maketitle

\begin{abstract}
There exists an unequivocal distinction between the sound produced by a static source and that produced by a moving one, especially when the source moves towards or away from the microphone. In this paper, we propose to use this connection between audio and visual dynamics for solving two challenging tasks simultaneously, namely: (i) separating audio sources from a mixture using visual cues, and (ii) predicting the 3D visual motion of a sounding source using its separated audio. Towards this end, we present \fullname (\name) -- a deep learning framework that leverages the 3D structure of the scene and the motion of sound sources for better audio source separation. At the heart of \name is a 2.5D scene graph capturing various objects in the video and their pseudo-3D spatial proximities. This graph is constructed by registering together 2.5D monocular depth predictions from the 2D video frames and associating the 2.5D scene regions with the outputs of an object detector applied on those frames. The \name task is then mathematically modeled as the joint problem of: (i) recursively segmenting the 2.5D scene graph into several sub-graphs, each associated with a constituent sound in the input audio mixture (which is then separated) and (ii) predicting the 3D motions of the corresponding sound sources from the separated audio. To empirically evaluate \name, we present experiments on two challenging audio-visual datasets, viz. Audio Separation in the Wild (ASIW) and Audio Visual Event (AVE). Our results demonstrate that \name achieves a clear improvement in source separation quality, outperforming prior works on both datasets, while also estimating the direction of motion of the sound sources better than other methods.
\end{abstract}
\section{Introduction}

Events around us are often audio-visual in nature and our senses have evolved to leverage this multimodal synergy to better reason about the world. For example, the sight of a kid laughing aloud while sliding down a slide allows us to associate \emph{the laughing sound} with the kid and get a sense of his/her direction of motion, even when a myriad of other sounds are present in the scene. In this paper, we propose to leverage this synergy between sight and sound for solving two challenging tasks simultaneously, viz.: (i) separating audio sources from a mixture using visual cues, and (ii) the novel task of predicting the 3D visual motion of the sounding source using its corresponding separated audio.

Typical approaches to visually-guided audio source separation use weakly- or self- supervised models trained to separate a mixture of acoustic sources into its constituents by conditioning on appropriate visual regions ~\cite{chatterjee2021visual,gao2019co,zhao2019sound,zhao2018sound}. Such approaches impose constraints on the space of the separated audio (e.g., cyclic consistency, object identifiability, etc.) to derive gradients for training the underlying neural models. A few of these methods additionally seek to ground the separated audio with the appearance of the sounding object~\cite{chatterjee2021visual,gao2019co,tian2021cyclic}. Chatterjee \etal~\cite{chatterjee2021visual}, construct a 2D scene graph on the video frames to capture the visual context of the audio source for better separation; their key intuition being that certain sounding objects (e.g., a guitar) cannot produce the sound by itself, but must have a suitable spatial context around them (e.g., a person). However, all the above approaches ignore one key aspect of the physical world -- that it is three dimensional and this 3D structure influences the sound being heard. For example, if \emph{a person playing a guitar} is spatially distant from the microphone, then the audio mixture to be separated is unlikely to have the sound of that guitar. 
Further, the 3D scene structure also allows for incorporating the motion of sounding objects. Imagine \emph{the whistling of a train moving towards you}. There is an inevitable relationship between the evolution of the sound of the train being heard and its 3D motion, which can  help distinctly separate its sound from other whistling trains or background clamour and vice-versa. Thus, if the audio of this train is well-separated via visual-guidance, then the separated audio must also be able to predict the direction of the train's motion. Leveraging these insights, we present \fullname (\name) -- an innovative graph neural network for video-guided audio source separation from an acoustic mixture that can also predict the direction of motion of the sound source.

Inspired by Cherian et al.~\cite{cherian20222}, our \name framework begins by computing a dense 2.5D representation of the frames of a video where 2.5D refers to the 2D visual context of the frames enriched with the pseudo-depth for that frame produced using a 2D-to-3D monocular depth prediction method~\cite{ranftl2021vision}. Next, we succinctly capture the semantic context of this 2.5D visual scene by means of a novel 2.5D scene graph representation. The nodes of this scene graph capture the various objects in the scene (projected onto the pseudo-3D scene via the depth map), while the graph edges characterize the approximate 3D spatial distances between the objects. Note that our graph does not explicitly capture any spatio-temporal dynamics of the scene objects, instead is constructed on singleton frames. To achieve audio source separation, we propose a recurrent graph neural network that is trained to segment this 2.5D scene graph into sub-graph embeddings; each of which is trained to be associated with a potentially unique \emph{sounding object or interaction} and is used to induce separation of that sound source from the audio mixture. During training, we enforce this uniqueness via imposing orthogonality constraints between the generated sub-graph embeddings. To make \name associate the evolution of sound with the 3D motions of their sources, we propose to include an auxiliary task that demands the prediction of the 3D direction of motion of a sounding object from its separated audio, where the ground truth 3D motion is estimated from the 2.5D scene graph using optical flow. 
 
 A natural question one may have at this point is: \emph{how can a method predict the 3D motion from monaural audio using a 2.5D scene graph constructed on a single video frame?} The key insights come from two observations: (i) there are often implicit object motion cues present in singleton video frames, which may be embedded within the scene graph node features (e.g., a graph node corresponding to a \emph{train} may have the inductive priors suggesting the train is moving in the direction it faces), and (ii) when the audio is separated via visual-guidance, these implicit features modulate the audio separation masks, thereby incorporating motion cues into the separated audio spectrograms. Thus, when trained end-to-end for the joint task of sound separation and motion prediction, the model leverages these implicit cues to minimize the learning objective.

 To demonstrate the efficacy of \name on sound source separation and motion direction prediction, we present experiments on two challenging datasets, namely (i) Audio Separation in the Wild (ASIW)~\cite{chatterjee2021visual} and (ii) Audio Visual Event (AVE)~\cite{tian2018audio}. Both of these datasets feature videos of sounding objects and sounding interactions \emph{in the wild}. Our results clearly show that \name outperforms competing prior methods on these datasets on both the tasks, underscoring the importance of incorporating 3D spatial structure and the benefits of learning audio-visual dynamics.

 \begin{figure}[t]
    \centering
    \includegraphics[width=0.9\linewidth]{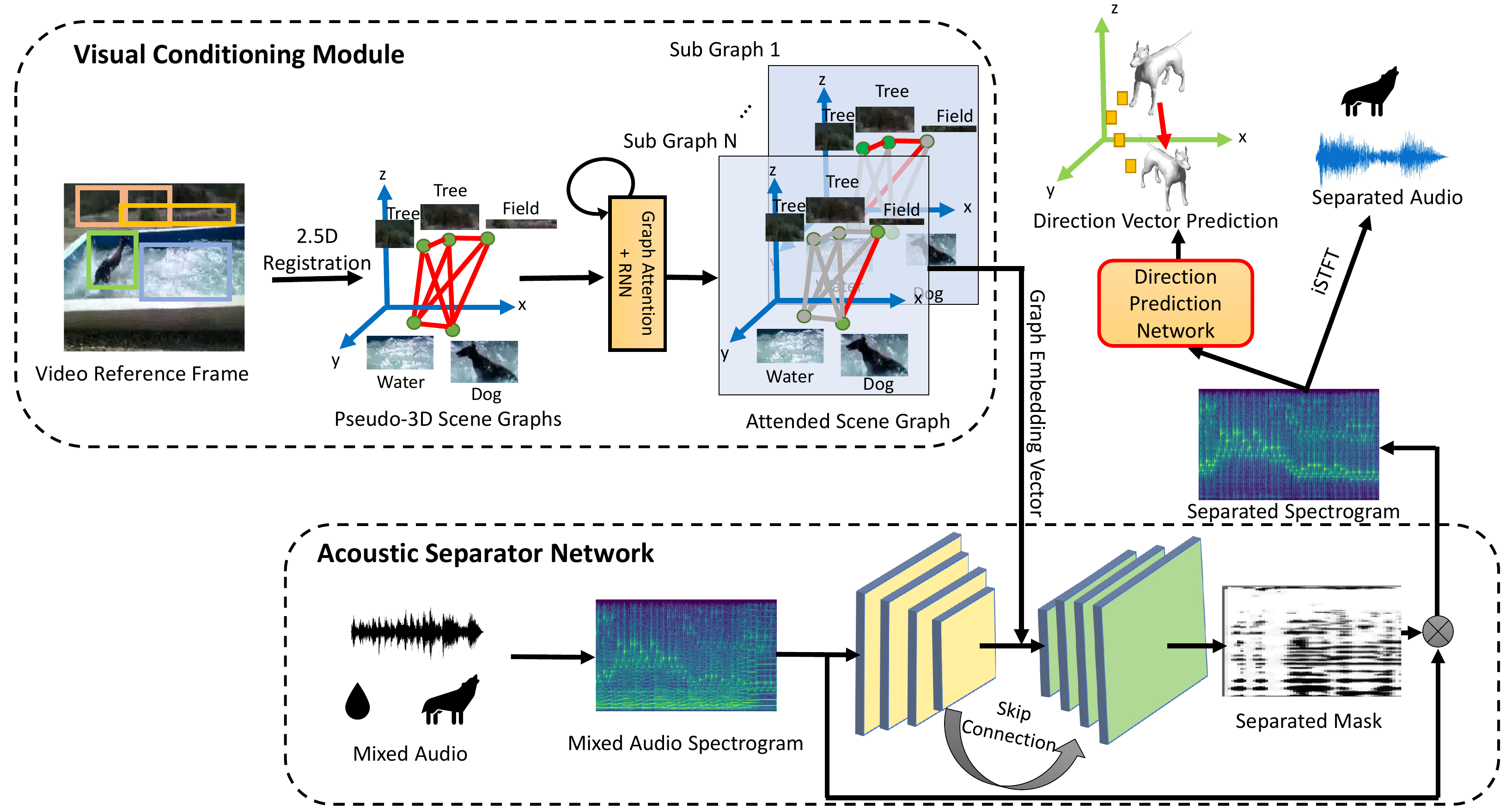} 
    
   \caption{A detailed illustration of our proposed \name model. }
    \label{fig:arch}
\end{figure}

Below, we summarize the key contributions of the paper:
\begin{itemize}\setlength\itemsep{-0.2em}
    \item We introduce a novel \emph{3D geometry-aware scene graph representation}~\cite{johnson2015image} for visually-guided audio source separation, called \name. 
	
	\item We introduce a novel task of \emph{predicting 3D motion direction} of a sound source in the scene from the the temporal evolution of the sound it makes, aided by appropriate visual context, and use it to improve audio separation.

	\item \name demonstrates \emph{state-of-the-art} audio source separation and motion prediction performances on two challenging datasets for this task, viz. ASIW and AVE.
\end{itemize}

\section{Related Works}

Below, we briefly review some of the important prior works  closely related to our approach.

\noindent \textbf{Visually-guided Audio Source Separation} is the task of separating a mixture of audio signals into its constituent sources by discovering the association between the separated sound sources and their visual appearances in the physical world. Typical methods for solving this task find a wide range of real-world applications, such as separating sounds of musical instruments~\cite{gao2019co,chatterjee2021visual,gan2020music,zhao2019sound,zhao2018sound}, separating speech signals~\cite{Afouras20b,ephrat2018looking,michelsanti2020overview}, and separating on-screen sounds from off-screen ones for generic objects~\cite{owens2018audio,tzinis2020into}. A learning pipeline often used to train such methods conditions a sound-separation network (such as a U-net~\cite{ronneberger2015u}) with an appropriate visual object representation, to induce a source separation on a mixed audio input, obtained by mixing the audio streams of two different videos. In recent years, research in this area has moved from using global visual features of motion and appearance~\cite{zhao2019sound} towards extracting very fine-grained visual conditioning information to induce more effective separations~\cite{gao2019co,tian2021cyclic}. Additional refinement steps have also been explored~\cite{chatterjee2021visual} to ward off potential false separation triggers. However, none of these approaches factor in the 3D geometry of the scene captured by the video -- an important gap that we attempt to fill using \name. 

\noindent \textbf{Localizing Sound in Video Frames} has been attempted in several recent methods by learning to ground the sound source in the visual space, e.g., identifying the pixels of the sound source~\cite{arandjelovic2018objects,hershey1999audio,kidron2005pixels,senocak2018learning,tian2021cyclic}. While, these approaches strive to learn the association between the visual appearance and acoustic signatures of sound sources, they do not apply such methods to audio source separation -- a task that is the focus of this work. 

\noindent \textbf{Sound Synthesis from Videos} constitutes another category of important techniques in the audio-visual realm~\cite{owens2016visually,zhou2018visual} that has become popular lately. Towards this end, several approaches have recently been proposed for the task of generating both monaural and binaural audio starting from videos~\cite{gao20192,morgado2018self,xu2021visually,li2021binaural}. However, differently, we seek to solve the task of sound source separation and motion prediction in the visual space.

\noindent \textbf{Application of Scene Graphs to Videos} has resulted in massive strides in video understanding tasks. Scene graphs, while traditionally used for capturing the static content of images~\cite{johnson2015image,li2008modeling} have lately been used for several video understanding tasks. For instance, Ji \etal~\cite{ji2020action} applied them for action recognition, Geng \etal~\cite{geng2020spatio} for visual dialog, and Chatterjee \etal~\cite{chatterjee2021visual} for visually-guided sound-source separation. However, these scene graphs are usually 2D, while we attempt to explicitly incorporate the 3D scene geometry into the scene graphs. While, Cherian et al.~\cite{cherian20222} proposes (2.5+1)D scene graphs for video question answering, our task of audio source separation and motion prediction brings in several novel components beyond their setup.

\noindent\textbf{Audio Separation Using Direction of Arrival} has been an important topic of recent interest in the audio research community~\cite{xu2021multiple,luo2019fasnet,ochiai2020beam,subramanian2021directional}, where the direction of arrival of sound to a microphone array is explicitly used for improved sound source separation. While, our approach is inspired by their key findings, we attempt to explore this in the audio-visual domain.

\section{Proposed Method}

\subsection{Task Setup and Method Overview}

Let $\vid$ denote an unlabeled video, and $m(t) = \sum_{i=1}^N\src_{i}(t)$ be its accompanying discrete-time audio arising from a linear mixture of $N$ acoustic sources $\src_{i}(t)$. The main goal of \textit{visually-guided audio source separation} is to use cues derived from $\vid$ to induce a separation of $m(t)$ into its constituent acoustic sources $\src_{i}(t)$, for $i \in \{1,2, \dots, N\}$. For effectively capturing the audio-visual semantic alignment, we represent the video $\vid$ as a scene graph $\graph = (\nodes, \edges)$ with nodes $\nodes = \set{\node_1, \node_2, \dots, \node_K}$ denoting the objects in $\vid$ and $\edges$ capturing the set of edges $\edge_{ij}$ characterizing the spatial proximity of a pair of nodes $(\node_i, \node_j)$. The key idea in our proposed \fullname (\name) framework is to map each of the acoustic sources $\src_{i}(t)$ to a sub-graph of $\graph$. We realize this mapping by auto-regressively partitioning the graph $\graph$ into mutually-orthogonal neural embeddings and using these embeddings to condition an \emph{Acoustic Separator} sub-network tasked with extracting a sound source $\src_{i}(t)$ from the mixture $m(t)$. A key ingredient in our model is equipping the audio separator to also predict the direction of motion of the sounding object. We incorporate this ability by including a direction prediction training loss. Figure~\ref{fig:arch} provides an overview of our model and the details follow.

 \begin{figure}[t]
    \centering
    \includegraphics[width=0.9\textwidth]{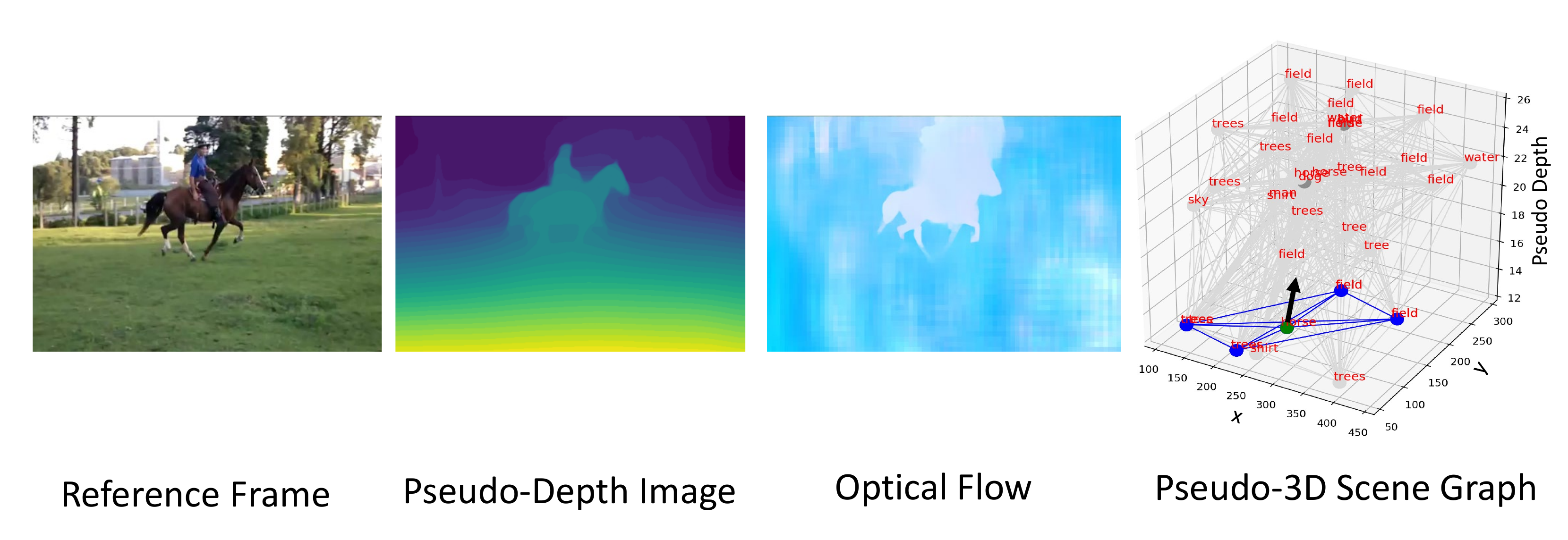} 
   \caption{A step-wise illustration of the pseudo-3D scene graph construction process on an example frame from an ASIW video. We show the
  original (reference) frame, its predicted monocular 3D depth map, its optical flow estimate with an adjacent frame, and a pseudo-3D scene graph for the frame. The highlighted part of the graph (blue edges) is the sub-graph associated with the sound of the horse's motion, while the black-arrow denotes the motion direction of the horse in pseudo-3D space. Note that the motion direction is not used in our graph embeddings, instead provides ground truth guidance when minimizing our losses.}
    \label{fig:asiw_dirpred}
\end{figure}

\subsection{Estimating Ground Truth Sound Source Motion}

One novel aspect of \name is that it can predict the motion direction of a source from its sound, which in turn is derived from an embedding of a sub-graph of $\graph$. However, the ground truth for this task is not directly available from the 2D frames of a given video. In the following, we detail the pre-processing steps for estimating this ground truth, and present an overview of these steps in Figure~\ref{fig:asiw_dirpred}. We denote a motion displacement vector, as $\disp_{i}$ for the source $\src_{i}(t)$, which will be subsequently used as an auxiliary cue to train the audio source separator.

\noindent \textbf{Computing Pseudo-Depth Maps}: Constructing a 3D scene from 2D images is a classic problem explored in computer vision for which a variety of solutions exist~\cite{agarwal2011building,hartley2003multiple}. However, many of these methods make strong assumptions, such as: (i) the scene being static, (ii) existence of sufficient overlap between the frames, or (iii) knowledge of the intrinsic parameters of the camera, none of which might hold for the type of videos that are typically used in our task. To this end, we propose to build upon the recent progress made in monocular 3D reconstruction realm towards mapping each video frame $\frm_f$ of $\vid$ into a 2.5D pseudo-depth image~\cite{zhao2020monocular}. In particular, we use the MiDAS algorithm~\cite{ranftl2021vision} because of its ease-of-use and accuracy of depth predictions.

\noindent \textbf{Rectifying Frames for Camera Motion:} Since we do not assume that the frames in $\vid$ have been captured by a static camera, we next align them to a common 3D reference frame $\frm_r$. To this end, we apply the classic \emph{Lucas-Kanade} tracker~\cite{lucas1981iterative} across the frames. As there may be objects moving out of the scene or shot-switches in the video, we found that tracking the \emph{static objects} in the video typically does not sustain its full length. Thus, we group the frames into fixed-size windows, each with a maximum of $\frmwind$ continuous frames. A video can therefore, be considered as a collection of $W$ windows, with each window producing its own motion displacement vector. This consequently also implies that our frames are aligned (rectified) within their respective windows. Concretely, let $\point$ denote a tracked point, in pseudo 3D, within a window. Suppose $\point_f\in\real^{3}$ and $\point_r\in\real^{3}$ be the 3D coordinate corresponding to $\point$ in $\frm_f$ and that in the reference frame $\frm_r$, respectively. Then, using the full set of such tracked points in this window, we rectify $\frm_f$ to $\frm_r$ by estimating a 3D transformation matrix, accurate upto a scale, using the iterative closest point (ICP) algorithm~\cite{besl1992method}. Once the target frame $\frm_f$ has been rectified, we also rectify its corresponding pseudo-depth map using the same transformation matrix.

\noindent \textbf{Visual Detection of Sound Sources:} To construct our scene graph, we first use a Faster-RCNN ($\frcnn$)~\cite{ren2016faster} model for detecting a set of $M$ objects and their 2D spatial bounding boxes in the reference frame $\frm_r$ of a window. The detection model is trained (on the Visual Genome dataset~\cite{krishna2017visual}) to recognize and localize 1600 commonplace object classes. However, some of the data that we use in our experimental study consists of sound producing objects that are not in the Visual Genome classes list. To accommodate such classes (e.g., musical instruments), we use a separate $\frcnn$ model pre-trained on images annotated with these classes in the Open Images dataset~\cite{krasin2017openimages}. Let $\objclasses$ denote the set of all detectable classes by either of these two detectors. For a frame $\frm_r$, the $\frcnn$ model produces a set of $M$ quadruples: $\set{\left(C^{i}_{r}, B^{i}_{r}, \feat^{i}_{r}, S^{i}_{r}\right)}_{i=1}^M = \frcnn(\frm_r)$, one for each detected object, consisting of the label  $C\in\objclasses$ of the detected object, its bounding box $B$ in the frame, a feature vector $F$ characterizing the visual appearance of the object, and a detection confidence $S$. 

\noindent \textbf{Tracking a Sound Source:} In order to robustly track the visual trail of a potential sound source between a reference frame $\frm_r$ and a target frame $\frm_f$, we employ RAFT~\cite{teed2020raft} -- an off-the-shelf state-of-the-art robust optical flow method. RAFT yields a dense optical flow field for each pixel in the reference frame. We filter this flow field using the bounding boxes $B^{i}_{r}$, and associate it with the $i$-th detected sounding object. Next, we base the flow field in a 3D coordinate space using the estimated pseudo-depth map associated with the respective frames; thus allowing for the recovery of the motion track of the sounding object in a pseudo-3D space.

\noindent \textbf{Estimating the Ground-Truth Displacement Vector:} Our proposed flow-based tracking method yields numerous displacement vectors (one for each pixel) within a detected object bounding box, this box corresponding to a potential sound source in the reference frame $\frm_r$. However, as alluded to above, we envisage representing a scene succinctly using a scene graph with each node capturing a single motion vector. To this end, we summarize the flow field associated with a detection box using its median; thus charaterizing the 3D displacement $\disp_i$ of that object in that window.

\subsection{Constructing the Pseudo-3D Visual Scene Graph}
With the object detections in place, as described above, we now proceed to construct the \textit{Pseudo-3D Visual Scene Graph} for $\vid$. While, for most of the prior approaches~\cite{geng2020spatio,ji2020action}, all detected objects are trivially added to the graph as nodes, our use-case of sound source separation dictates that the constructed graph permits the learning of correlations between appropriate subgraphs and the separated audio sources. Towards this end, each sound source in a dataset is associated with a set of objects in the visual domain (i.e., from the set of classes $\objclasses$ of $\frcnn$) which we call \emph{Auditory Objects} and  which could potentially create sound. For instance, a \textit{crying} sound could arise from classes such as \emph{kid}, \emph{baby}, etc. Let us denote an auditory object by $\pobj$, where $\pobj \in \objclasses$ . The entire video $\vid$ could then be thought of as a collection of auditory objects, $\pobjs= \{\pobj_{1}, \dots, \pobj_{{N}}\}$, which together generate the mixed audio $m(t)$. For each of these auditory objects $\pobj_{i}$, we identify the subset $\nodes_{\pobj_{i}}$ of $M - N$ objects that are non-auditory and overlap with $\pobj_{i}$ with an Intersection Over Union (IoU) greater than a pre-defined threshold $\gamma$. These objects are called \emph{Context Objects}. The vertex set $\nodes$ of the graph $\graph$ associated with the video $\vid$ is then constructed as $\nodes = \bigcup_{i=1}^{N} (\set{\pobj_{i}} \bigcup  \nodes_{\pobj_{i}})$. 

Next, we describe the process of constructing the edge set $\edges$ of the graph $\graph$. While prior approaches that use scene-graphs, typically either design dense fully-connected scene-graphs with equal edge weights~\cite{chatterjee2021visual} or deploy a visual relationship detector~\cite{geng2020spatio}, nonetheless they do not usually embody any 3D scene geometry. Instead in \name, we propose to mitigate this shortcoming by computing the edge weights from the pseudo-3D scene we constructed above. To this end, we first extract 3D point clouds for every object $i$ in frame $\frm_r$ by leveraging the rectified pseudo-depth map. Let $\mathit{Q}_i$ (for $i=1, 2, \dots M$) denote the pseudo 3D point cloud of all pixels associated with node $\node_i$  within its corresponding $\frcnn$ bounding box. Note that each point $\point_r^{\mathit{Q}_i} \in \mathit{Q}_i$ is a 3D vector with pixel-normalized $\mathbf{x}\mathbf{y}$ coordinates and their pseudo-depths. To define the spatial proximity between two nodes $\node_i$ and $\node_j$, we compute the symmetrized Chamfer distance ($D_{ij}$)~\cite{barrow1977parametric} between every pair of points in $\point_r^{\mathit{Q}_i}$ and $\point_r^{\mathit{Q}_j}$. 
These distances are then normalized to be within $[0,1]$ by dividing by  the maximum value over all edges in $\edges$ and are used to create a weighted-adjacency matrix defining the graph edges $\edges$ by using a radial basis function on the Chamfer distance matrix, i.e., the weight of an edge $\edge_{ij}=\exp (\frac{-D_{ij}}{\sigma^2})$, for a suitable scale $\sigma$, set in our experiments as the $25^{th}$ percentile of the entries in the distance matrix.

\noindent \textbf{Visual Encoding of Sounding Interactions:} The scene graph $\graph$ is a spatio-semantic summary of the visual information in $\vid$ and is the visual analogue to the mixed audio $m(t)$ accompanying $\vid$. For the purposes of our task, we are interested in extracting appropriate subgraphs of $\graph$ that are capable of inducing a separation of $m(t)$ into its constituents $\src_i(t)$ for $i = \set{1, 2, \dots, N}$. This is accomplished by employing a recurrent graph attention to segment the graph $\graph$ into sub-graphs. To segment the graph $\graph$, we first employ a multi-headed graph attention network~\cite{velivckovic2017graph} which operates on the features associated with  the scene graph nodes  and performs multi-head graph message passing. This allows \name to incorporate the spatial scene geometry (from edge weights) and scene context from the neighbors of an auditory node for characterizing the visual representation of a sound source. For capturing the visual representation of interactions (between objects) that produce a sound, we design an edge convolution network~\cite{wang2019dynamic}, the features from which are then simultaneously pooled with the graph attention features using global max-pooling and global average pooling~\cite{lee2019self} and concatenated. This results in an embedding vector $\vzeta$ for the entire graph. Since we need to extract separate sub-graph embeddings to induce a separation of the sound mixture into $N$ sources, we employ a gated recurrent unit (GRU), to recursively extract visual feature vectors $\embedset=\set{\vy_1, \vy_2,\dots, \vy_N}$ from $\vzeta$. We noticed that extracting a separate background sound from every mixture helps. Thus, our GRU is executed for $N+1$ steps, rather than $N$.

To ensure that the extracted graph embeddings $\embedset$ from the GRU do not repeat a sub-graph, we enforce mutual orthogonality~\cite{chatterjee2021visual} between these embeddings. That is, each embedding $\vy_i$ emerging out of a recurrence of the GRU is expected to produce a unit-normalized embedding $\hat{\embed}_i$ that is orthogonal to each of the unit-normalized embeddings generated prior to it, i.e., $\set{\hat{\embed}_1, \hat{\embed_2},\dots, \hat{\embed}_{i-1}}$. This constraint is incorporated as a regularization in our training setup. Mathematically, a softer-version of this constraint given by the following is implemented:
\begin{align}
    \loss_{\mathrm{ortho}}(\embedset) = \sum_{i, j \in \{1,2, \dots, N\},  i \neq j} (\hat{\embed}_i^\top {\hat{\embed}}_j) ^{2}.
    \label{eq:dotpdt}
\end{align}

\noindent \textbf{Acoustic Separator Network:} Finally, to induce the separation in the acoustic signal, we incorporate an \textit{Acoustic Separator Network} (ASN), which is a U-Net \cite{ronneberger2015u} style encoder-decoder network. Such networks have recently shown promise in sound source separation tasks~\cite{jansson2017singing,liu2019divide}, particularly those that are employed in conditioned settings~\cite{gao2019co,meseguer2019conditioned,slizovskaia2019end,zhao2018sound}. The network has three main parts: (i) an \emph{encoder} consisting of a stack of 2D-convolution layers, each coupled with Batch Normalization and Leaky ReLU, (ii) a \emph{bottleneck} layer that concatenates the encoder embedding with the conditioning information, and (iii) a \emph{decoder} which is made up of a series of up-convolution layers, followed by non-linear activations, each coupled with a skip connection from a corresponding layer in the U-Net encoder,  matching in spatial resolution of its output. The encoder encodes the input, which is a magnitude spectrogram ${\mspec}\in\reals{\Omega\times T}$ of a mixed audio $m(t)$ produced via the short-time Fourier transform (STFT), where $\Omega$ and $T$ denote the number of frequency bins and the number of video frames, respectively. This encoding is then concatenated with each of the normalized visual encoding vectors $\hat{\embed}_i$ and decoded to generate a time-frequency mask, $\hat{\mask}_{i} \in [0, 1]^{\Omega \times T}$, which when multiplied with the magnitude spectrogram $\mspec$ of the mixture yields an estimate of the magnitude spectrogram of the separated source $\hat{\srcspec}_{i}= \hat{\mask}_{i} \odot {\mspec}$, where $\odot$ denotes element-wise product. The separated signal in time domain, $\hat{\src}_{i}(t)$, is then recoverable by applying inverse Short time Fourier Transform (iSTFT).

\noindent \textbf{Direction Prediction Network:} A single-channel audio typically only indicates whether the sound source is moving towards/farther away from the camera~\cite{andrade1959doppler}. However the sub-graph embeddings, $\vy_i$, that the ASN is conditioned on, injects into the separated spectrograms, motion information implicit in the appearance of the objects in the frame. This enriches these spectrograms, permitting us to obtain a coarse estimate of the direction of source motion in pseudo 3D. 
We therefore cast the task of predicting the motion vector $\disp_{i}$ of the auditory object as a classification task. We setup two variants of this task using varied quantizations of the unit displacement vector $\disp_i$, namely: (i) to predict the octant (or corners) of a unit-cube into which an 8-quantized  unit displacement vector $\disp_{i}$ belongs, where the center of the cube refers to the 3D centroid of the corresponding graph node; and (ii) to predict among the 26 directions pertaining to the 8 corners, 6 faces, and 12 edges of a unit-cube,  $\disp_{i}$ is quantized into (using cosine similarity). Additionally, for both settings a class is created for denoting little/no motion and another class captures the direction vector obtained from the background source. We thus have 10 and 28-classes respectively, in the two settings. We realize this objective in \name by incorporating a classifier network $\vh_{\Lambda}(\cdot)$, with parameters $\Lambda$, that takes as input the (spectrogram) output of the ASN and produces the class label corresponding to the quantization of $\disp_{i}$. 

Note that we compute a separate displacement vector $\disp_{i}$ per window $w$, while the separated $\hat{\srcspec}_{i}$ spans the whole length of the video. We thus compute a set of displacement vectors $\set{\disp_{i}^{w}}_{w=1}^{W}$, for each window $w$. Our network takes as input the separated spectrogram, $\hat{\srcspec}_{i}$, and the visual encoding of the source $i$ ($\hat{\embed}_i$) and produces as output an estimate of the direction class, $\hat{\disp}_{i}^{w} \in \set{0, 1, 2, \dots, 9}$ or $\hat{\disp}_{i}^{w} \in \set{0, 1, 2, \dots, 27}$, depending on the direction in which the vector lies. In order to predict the vector for a particular window, appropriate time-aligned slices of $\hat{\srcspec}_{i}$ are chosen. The sliced spectrogram is then embedded through a ResNet~\cite{he2016deep} style network. 
Thus, we have $\vh_{\Lambda}:\hat{\srcspec}_{i} \rightarrow \set{\hat{\disp}_{i}^{w}}_{w=1}^{W}$, where the prediction for each window is treated as a different sample in the batch. Details in the appendix.

\subsection{Learning Losses}
\name trains purely with weak/self-supervisory cues. In particular, we train our model by employing the ``mix-and-separate'' strategy~\cite{chatterjee2021visual,gan2020music,gao2019co,zhao2019sound,zhao2018sound}, where the audio tracks from multiple videos (typically two) are combined, with the goal being to separate the audio for each possible source from the mixture ($\mspec$). Towards this end, we employ the following set of losses: 

\noindent \textbf{Consistency Loss} ensures that the separated spectrograms of a particular class of sound source (such as a sound from a guitar, the cry of a baby, etc.) always retains its (visual) identity irrespective of which video it comes from~\cite{gao2019co}. Typically a cross-entropy loss satisfies this requirement. However in \name, since the separation is induced by a holistic scene graph recurrently embedded by a GRU, it is not known in advance, in what order the sources are separated. Thus, we consider all possible combinations of the ground truth labels and assign the one with the minimum loss. This is given by:
\begin{equation}
\mathcal{L}_{\mathrm{cons}} =  \sum_{u=1,2} \min_{\sigma^u\in \mathcal{S}_{N_u+1}} -
\sum_{i=1}^{N_u+1} \sum_{c=1}^{K} \mathbbm{1}^u_{i, \sigma^u(c)}  \log p^u_{i, c},
\label{eq:consistency}
\end{equation}
where $N_u+1$ is the number of auditory objects in the $u$-th video,  $\mathcal{S}_{N_u+1}$ indicates the set of all permutations on $\{1,\dots,N_u+1\}$, $p^u_{i, c}$ denotes the predicted probability produced by the classifier for class $c$ given $\hat{\srcspec}^{u}_{i}$ as input, $\mathbbm{1}^u_{i, \sigma^u(c)}$ is an indicator for the ground-truth class of the $c$-th object in video $u$, and $K$ denotes the number of sounding object classes in the dataset.

\noindent \textbf{Cyclic Loss} attempts to preserve the consistency of the generated audio masks from the same video. Specifically, this loss ensures that the separated masks corresponding to each of the sources in the video, when combined, yields a mask which separates the audio spectrogram for that video from the mixture audio (that combines audio from two videos). Mathematically, our cyclic loss is written as: $\mathcal{L}_{\mathrm{cyc}} = \sum_{u=1,2} \Big\|\sum_{i=1}^{N_u + 1} \hat{\mask}^{u}_i - \mask^u_{\mathrm{ibm}}\Big\|_1$, where $\mask^u_{\mathrm{ibm}}=\mathbbm{1}_{\mspec^u > \mspec^{\neg u}}$ denotes the binary mask for the audio of video $u$ within the mixture $\mspec$.

\noindent \textbf{Direction Prediction Loss} is a standard cross-entropy loss, which checks for the accuracy of the predicted direction class $\set{0, 1, \dots, D_k}$, where $D_k\in {9, 27}$ for every separated sound source, per window. It is given by:
\begin{equation}
\mathcal{L}_{\mathrm{dirpred}} = \sum_{u=1,2}\sum_{w=1}^{W} \min_{\sigma^u\in \mathcal{S}_{N_u+1}} -\sum_{i=1}^{N_u+1}  \sum_{c=1}^{D_k} \mathbbm{1}^{u, w}_{i, \sigma^u(c)}  \log q^{u, w}_{i, c},
\label{eq:dirpred}
\end{equation}
where $N_u+1$ is the number of auditory objects in the $u^{th}$ video. $\mathcal{S}_{N_u+1}$ indicates the set of all permutations on $\{1,\dots,N_u+1\}$, $q^{u, w}_{i, c}$ denotes the predicted probability produced by the classifier for class $c$ given $\hat{\srcspec}^{u, w}_{i}$ as input, and $\mathbbm{1}^{u, w}_{i, \sigma^u(c)}$ is an indicator of the ground-truth class of the $c$-th object in video $u$ for window $w$.

\noindent\textbf{Final Training Loss} of \name is given by the following, with weights $\lambda_1, \lambda_2, \lambda_3, \lambda_4 > 0$:
\begin{equation}
\mathcal{L} = \lambda_1 \mathcal{L}_{\mathrm{cons}} + \lambda_2 \mathcal{L}_{\mathrm{cyc}} + \lambda_3 \mathcal{L}_{\mathrm{ortho}} + \lambda_4 \mathcal{L}_{\mathrm{dirpred}}.
\end{equation}

\section{Experiments}

\begin{table}[t]

 \centering
  \sisetup{table-format=2.1,round-mode=places,round-precision=1,table-number-alignment = center,detect-weight=true} 
 \caption{SDR, SIR, and SAR results on the ASIW and AVE test sets. 
 [Key: \textbf{Best}, \textcolor{blue}{second-best} results.] } 
 \resizebox{\linewidth}{!}{
 \begin{tabular}{lSSSSSSSS}
\toprule
\multirow{2}{*}{\textbf{$\qquad$ Approach}} & \multicolumn{3}{c}{\textbf{ASIW}} & \multicolumn{3}{c}{\textbf{AVE}} \\ 
\cmidrule(l{0.25em}r{0.25em}){2-4}\cmidrule(l{0.25em}r{0.25em}){5-7}
& {SDR } $\uparrow$ & {SIR } $\uparrow$ &	{SAR } $\uparrow$ &	{SDR } $\uparrow$ & {SIR } $\uparrow$ &	{SAR } $\uparrow$ \\ 
\midrule
Sound of Motion (SofM)~\cite{zhao2019sound} & 6.7	& 9.4 &	11.1 & 	4.1 & 9.21 &   7.62  \\  
Cyclic Co-Learn \cite{tian2021cyclic} & 7.00	& 13.39 &	12.41 &	4.16 &	9.66 & \color{blue}	8.42  \\ 
Co-Separation~\cite{gao2019co} & 6.6	& 12.9 &	12.6 & 	3.93 &	9.32 &	7.79  \\ 
AVSGS~\cite{chatterjee2021visual} & \color{blue} 8.8	& \color{blue}  14.1 & \color{blue}	13.0  & \color{blue} 5.77 & \color{blue}	10.35 &	8.15 \\ \midrule 

\name (only 2.5D graph) & \bfseries 9.02 &	\bfseries 14.32 & \bfseries	13.65 & 	\bfseries 6.52261214 & \bfseries	12.42221814 & \bfseries	8.89  \\ 

\name (2.5D graph + motion) & \bfseries 9.55 &	\bfseries 14.47 & \bfseries	14.12 &  	\bfseries 7.18 &\bfseries 13.32 	& 	\bfseries 9.43 \\ \midrule 
\midrule
\end{tabular}
}
\label{tab:perf_asiw_ave}

\end{table}

\begin{table}[t]
 \centering
 \sisetup{table-format=2.1,round-mode=places,round-precision=1,table-number-alignment = center,detect-weight=true} 
     \caption{Direction Prediction results on the ASIW and AVE on test splits.}
 \resizebox{1.0\linewidth}{!}{
 \begin{tabular}{lSSSS}
\toprule
\multirow{2}{*}{\textbf{Direction Prediction}} & \multicolumn{2}{c}{{ASIW}} & \multicolumn{2}{c}{{AVE}} \\ 
\cmidrule(l{0.25em}r{0.25em}){2-3} \cmidrule(l{0.25em}r{0.25em}){4-5}
 & \textbf{ 10-class (\%)}  $\uparrow$ & \textbf{ 28-class (\%)}  $\uparrow$ & \textbf{ 10-class (\%)}  $\uparrow$ & \textbf{ 28-class (\%)}  $\uparrow$\\ 
\midrule
Majority Vote & 27.29 & 25.36 & 29.17 & 24.27 \\
Sound of Motion (SofM)~\cite{zhao2019sound} & 29.62 & 26.98 &	31.23 & 30.55 \\ 
Cyclic Co-Learn \cite{tian2021cyclic} & 34.77 & 32.33 &	30.65 & 29.21 \\ 
Co-Separation~\cite{gao2019co} & 32.23 & 31.67 &	30.21 & 27.98 \\ 

AVSGS~\cite{chatterjee2021visual} & \color{blue} 39.15	& \color{blue} 38.65 &	\bfseries 38.92 & \color{blue} 34.65  \\ 
\midrule

\name (Ours) & \bfseries 42.51 &  \bfseries 41.31 &	\color{blue} 38.53 & \bfseries 36.84 \\ \midrule 
\midrule
\end{tabular}
}
\label{tab:main_dirpred}
\end{table}
\begin{table}[t]
 \centering
 \sisetup{table-format=2.1,round-mode=places,round-precision=1,table-number-alignment = center,detect-weight=true} 
 \caption{SDR, SIR, and SAR Ablation results on the ASIW test set. [Best results in \textbf{bold}.] }\label{tab:asiw_ablation}
 \resizebox{0.9\linewidth}{!}{
 \begin{tabular}{rlSSS}
\toprule
\multirow{2}{*}{} & & \multicolumn{3}{c}{{ASIW}} \\ \cmidrule(l{0.25em}r{0.25em}){3-5}
{Row}& {Method}&  {SDR  [dB]} $\uparrow$ & {SIR  [dB]} $\uparrow$ &	{SAR  [dB]} $\uparrow$  \\ 
\midrule
 1 & \name (Full Model) & \bfseries 9.55 &	\bfseries 14.47 & \bfseries	14.12  \\ \midrule
2 & \name - Multiscale Chamfer &  9.23 &	14.08 & 	13.97  \\ 
3 & \name - Only 10-class Direction Prediction ($\lambda_1 = \lambda_2 = \lambda_3 = 0$) &  6.4 &	11.2 &	11.7  \\ 

\bottomrule
\end{tabular}
}
\label{tab:ablation}
\end{table}

In this section, we demonstrate the effectiveness of our \name approach on challenging visually-guided audio separation benchmarks. 
\subsection{Datasets and Experimental Setup}
\noindent \textbf{Audio Separation in the Wild (ASIW) Dataset:} Several prior works present experiments mainly for separating sounds of musical instruments~\cite{gan2020music,tian2021cyclic}. Such results may not necessarily carry over easily to more generic and natural day-to-day life contexts. To this end, we consider the recently proposed \emph{Audio Separation in the Wild} (ASIW) dataset~\cite{chatterjee2021visual} for our experiments. This dataset consists of 147 validation, 322 test, and 10,540 training videos crawled from the larger AudioCaps dataset~\cite{kim2019audiocaps}. It has 14 auditory classes (i.e. $K=15$, 14 auditory classes and 1 background audio class), such as \emph{baby cries}, \emph{bell ring}, \emph{birds chirp}, \emph{camera clicks}, \emph{train sounds}, \emph{automobile sounds}, etc. Each video in the dataset is 10s long, has significant camera motion, and captures diverse audio-visual contexts. To evaluate \name for direction prediction, we divide each video into temporal windows of 1s long and predict a discretized 3D direction vector for a randomly chosen window. 

\noindent \textbf{Audio Visual Event (AVE) Dataset}: Apart from the challenging ASIW dataset, we conduct additional experiments by adapting the popular \emph{Audio Visual Event} (AVE) Dataset~\cite{tian2018audio} for our task. Since this dataset was originally designed for the task of identifying audio-visual events, adapting it to evaluate our approach is straightforward. We treat each audio-visual event class as an auditory class and associate with it a set of \emph{Auditory Objects} (as discussed before). The dataset contains 2211 training, 257 validation, and 261 test set videos. We use videos corresponding to 18 audio-visual event classes (i.e. $K=19$ here), including both potentially-moving object classes such, such as bus, train, etc. and static classes like banjo, clock etc. The videos in this dataset are 10s long as well. Similar to ASIW, we divide each video into windows of 1s long for computing the motion displacement vectors.

\noindent \textbf{Baselines: } To the best of our knowledge, the task of jointly separating audio sources (using video) and predicting motion directions is a novel task, and thus does not have a prior baseline to compare to. Therefore, we compare \name against the following state-of-the-art methods for audio-visual source separation, specifically those methods that use the ``mix-and-separate'' learning and also evaluate against them for assessing the efficacy of direction prediction. Specifically, we compare to:
(i) \textit{Co-Separation}~\cite{gao2019co} that uses a single auditory object for conditioning a source separation network,
(ii) \textit{Sound of Motion} (SofM)~\cite{zhao2019sound} that incorporates object/human appearances and their pixel-level motion trajectories to condition an audio separation network, 
(iii) \textit{Cyclic Co-Learning}~\cite{tian2021cyclic}, which is a recent method that trains a deep neural model for the joint task of audio separation and visual grounding of the audio source, and (iv)
\textit{Audio-Visual Scene Graph Segmenter} (AVSGS)~\cite{chatterjee2021visual}, that uses a scene graph for conditioning an audio separator network using the 2D visual context of the audio source.

\noindent \textbf{Evaluation Metrics: } We quantify the audio source separation performances using the standard evaluation metrics~\cite{gao2019co,tian2021cyclic,zhao2018sound}, namely: (i) \emph{Signal-to-Distortion Ratio (SDR)} (in dB) \cite{raffel2014mir_eval,Vincent2006BSSeval} -- a higher SDR indicating a more faithful reproduction of the original signal, (ii) \emph{Signal-to-Interference Ratio (SIR)} (in dB) -- quantifying the extent of reduction in interference in the estimated signal, and (iii) \emph{Signal-to-Artifact Ratio (SAR)} (in dB) -- capturing the extent to which artifacts are introduced by the separator. Further, as our motion direction prediction sub-task is cast as a classification problem in our setup, we also report its classification accuracy (Dir. Acc. ($\%$)) for both 10 and 28 class settings.

\noindent \textbf{Implementation Details: } For every video in our setup, we detect upto two auditory objects, one background auditory object, and 20 context nodes. The visual feature of the background object is derived from a $\frcnn$ embedding of a random crop of the reference frame $\frm_r$, from a region which does not overlap with other boxes in the frame. The audio streams are sub-sampled at 11kHz and STFT spectrograms are extracted using a Hann window of size 1022, with hop length of 256, following prior works~\cite{gao2019co,zhao2018sound}. We set $\Omega=256$ and $T=256$. The embeddings, $\hat{\embed}_i$, and the GRU-hidden state are $512$-dimensional. The IoU threshold, $\gamma$ is set to 0.1 for both datasets. Each window in a video has $l=8$ frames. The weights on the different losses are as follows: $\lambda_1=0.05, \lambda_2=1.0, \lambda_3=1.0, \lambda_4=0.05$. Our model is trained using the ADAM optimizer~\cite{kingma2014adam} with a weight decay of $1\text{e-}{4}$, $\beta_1=0.9$, $\beta_2=0.999$. The learning rate is set to $1e-4$ and is decreased by a factor of 0.1 every 15K iterations.

\begin{figure}[t]
    \centering
    \includegraphics[width=0.9\textwidth]{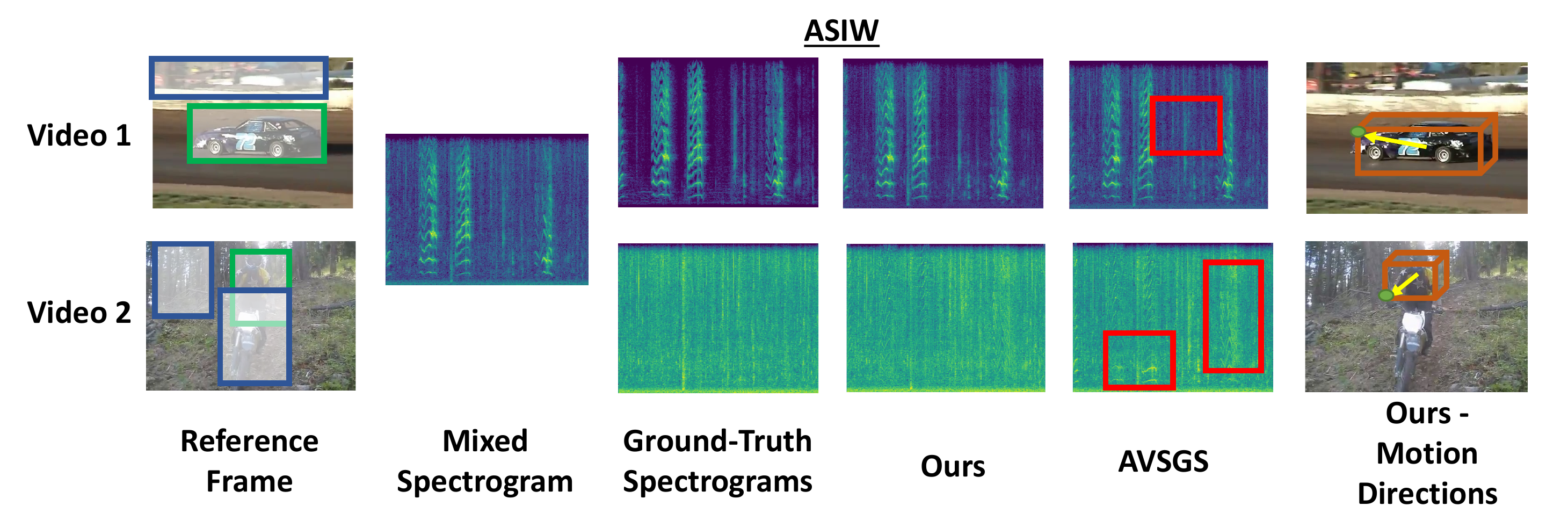}
    \includegraphics[width=.9\textwidth]{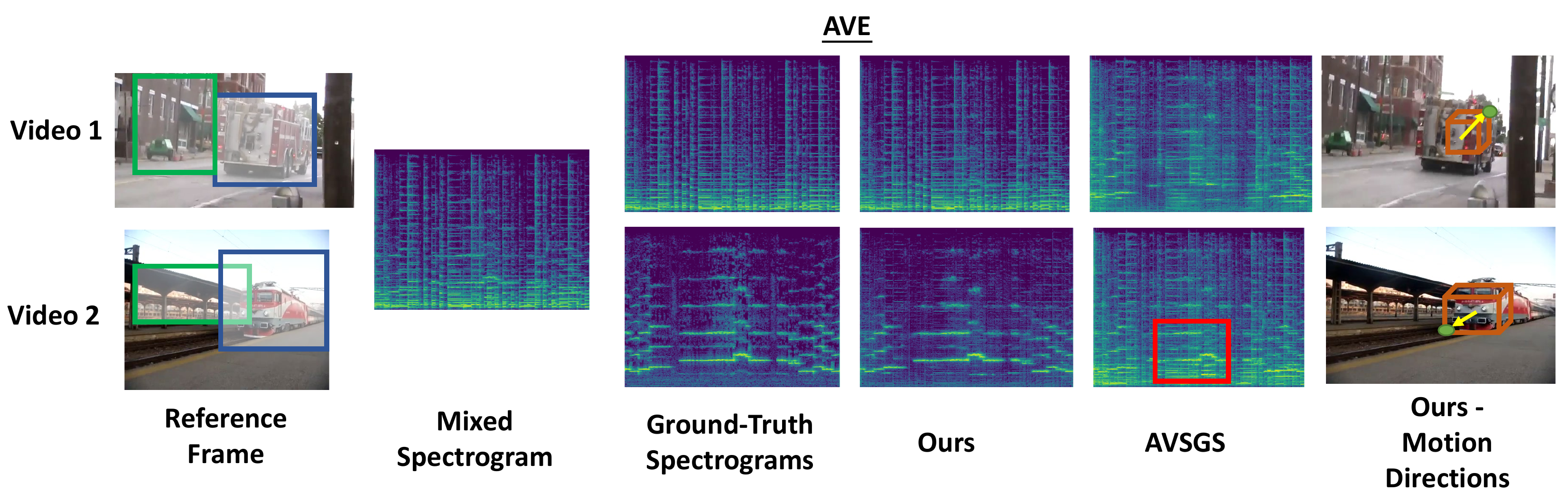}
   \caption{ Audio source separation and motion prediction results on ASIW (top) and AVE (bottom). \textbf{Bounding boxes} on reference frames show regions attended by \name. \textcolor{red}{Red} boxes indicate regions of high differences between ground truth and predicted spectrograms. The \textcolor{orange} {orange} unit cube shown in the \emph{Motion Directions} images, indicates the object whose motion is predicted. The \textcolor{green}{green} dot shows the ground truth motion direction, while the \textcolor{yellow}{yellow} arrow shows the predicted direction by \name.} 
    \label{fig:model_perf}
\end{figure}

\subsection{Experimental Results}
In Table~\ref{tab:perf_asiw_ave}, we present results for source separation by our method versus competing approaches on the ASIW and AVE datasets.  In Table~\ref{tab:main_dirpred}, we report the direction prediction accuracy of \name on the same set of competing methods for both 10 and 28 class setting. Our results clearly attest to the benefit of using the 3D structure of the scene and the motion direction prediction for audio separation. Specifically, \name itself (without motion prediction) yields a boost of upto 0.7 dB SDR over prior methods, just on source separation. Additionally, the performance of \name improves by atleast another 0.6 dB (on the SDR scale) across the two datasets, when trained to predict the motion direction (Table~\ref{tab:perf_asiw_ave}). Furthermore, the results evince that \name outperforms prior approaches in almost all of the evaluation categories, by upto 2.0 dB on the SIR scale on the AVE dataset over the next best (and closely-related) approach of AVSGS~\cite{chatterjee2021visual}. These results clearly demonstrate the generalizability of our method to varied datasets, scene contexts, and audio diversity. From amongst the competing techniques, we find that AVSGS comes closest to matching our performance across the two datasets, which is perhaps unsurprising given that they use a 2D scene graph to encode the visual context of the sounding object. However, our results show that the presence of 3D scene geometry is also essential for good audio separation (Table~\ref{tab:perf_asiw_ave} last two rows).

In Table~\ref{tab:main_dirpred}, we also report the accuracy of motion prediction of the sounding object (in \%) for both \name and other baselines for both 10-class and 28-class settings by passing their separated spectrograms through the direction prediction module. As is clear from the table, we find that the accuracy of direction prediction using \name is much better than prior methods (by upto $\sim 3\%$). The gains over a simple majority label prediction is especially pronounced, attesting to the efficacy of our direction prediction pipeline.

\noindent \textbf{Ablation Study: } %
We report performances of a few key ablated variants of our model in Table~\ref{tab:asiw_ablation} on the ASIW dataset, viz. : (i) by computing the weights of the graph edges via a multiscale RBF kernel, (ii) using only the motion direction prediction as a training signal. For the former, we threshold the Chamfer distances at two scales, viz : (a) the median of the weights and (b) $100^{th}$ percentile of the weights. This creates a forest of graphs, which is used in the audio separator network. Our results in Table~\ref{tab:asiw_ablation} show that this approach of graph construction levels up to our RBF formulation but falls a bit short, perhaps because of the hard thresholding of edges. In another ablation study, we train only the motion direction prediction loss (i.e., $\lambda_1=\lambda_2=\lambda_3=0$). As the results in Table~\ref{tab:asiw_ablation} show, this variant underperforms our full model, corroborating the utility of the other loss terms.

\noindent \textbf{Qualitative Results: } In Figure~\ref{fig:model_perf}, we present sample separation results from the ASIW and AVE test sets, juxtaposing the results obtained by our method against the AVSGS baseline. The superiority of our model's separability is evident even at a glance of the separated spectrograms that mimic the ground truth spectrograms more closely compared to AVSGS. The estimates of the motion directions seem to agree with the ground truth as well. Additional details and results are in the appendix. 

\section{Conclusions}
In this work, we presented \name, a novel algorithm to leverage visual scene geometry to induce better separation of a mixed audio signal into its semantic constituents. Towards this end, we incorporated pseudo-3D cues into our algorithmic audio separation pipeline via a scene graph data structure. Our work further derives self-supervision cues from the visual dynamics inherent in the videos and use such cues to predict the direction of motion of the sound source from the separated audio; our experiments demonstrate this self-supervision to improve source separation. \name demonstrates state-of-the-art performances on two challenging ``in-the-wild'' datasets. The success at the task of predicting the direction of motion from the visually-guided separated audio suggests \name's potential to be used in a variety of other audio-visual problems, including occlusion reasoning, and video generation from audio~\cite{chatterjee2020sound2sight}. We will be making our PyTorch implementation public at: \href{https://sites.google.com/site/metrosmiles/research/research-projects/asmp}{https://sites.google.com/site/metrosmiles/research/research-projects/asmp}

 \noindent \textbf{Limitations:} While \name presents promising results on both separation and direction prediction tasks, some caveats remain.  For example, our model is not yet capable of effectively inducing source separation for sounds emanating from the background. Further, we require trained object detectors for sound producing classes, which for some rare/uncommon classes maybe hard to obtain.

\noindent \textbf{Acknowledgements.} MC and NA would like to thank the support of the Office of Naval Research under grant N00014- 20-1-2444, and USDA National Institute of Food and Agriculture under grant 2020-67021-32799/1024178.

{
\small
\bibliographystyle{plain}
\bibliography{smp_neurips22}

\begin{thebibliography}{10}

\bibitem{Afouras20b}
Triantafyllos Afouras, Andrew Owens, Joon~Son Chung, and Andrew Zisserman.
\newblock Self-supervised learning of audio-visual objects from video.
\newblock In {\em Proc. ECCV}, 2020.

\bibitem{agarwal2011building}
Sameer Agarwal, Yasutaka Furukawa, Noah Snavely, Ian Simon, Brian Curless,
  Steven~M Seitz, and Richard Szeliski.
\newblock Building rome in a day.
\newblock {\em Communications of the ACM}, 54(10):105--112, 2011.

\bibitem{andrade1959doppler}
EN~da~C Andrade.
\newblock {\em Doppler and the Doppler effect}.
\newblock Imperial Chemical Industries Limited, London, 1959.

\bibitem{arandjelovic2018objects}
Relja Arandjelovic and Andrew Zisserman.
\newblock Objects that sound.
\newblock In {\em Proc. ECCV}, pages 435--451, 2018.

\bibitem{barrow1977parametric}
Harry~G Barrow, Jay~M Tenenbaum, Robert~C Bolles, and Helen~C Wolf.
\newblock Parametric correspondence and chamfer matching: Two new techniques
  for image matching.
\newblock Technical report, SRI INTERNATIONAL MENLO PARK CA ARTIFICIAL
  INTELLIGENCE CENTER, 1977.

\bibitem{besl1992method}
Paul~J Besl and Neil~D McKay.
\newblock Method for registration of 3-d shapes.
\newblock In {\em Sensor fusion IV: control paradigms and data structures},
  volume 1611, pages 586--606. Spie, 1992.

\bibitem{chatterjee2020sound2sight}
Moitreya Chatterjee and Anoop Cherian.
\newblock Sound2sight: Generating visual dynamics from sound and context.
\newblock In {\em Proc. ECCV}. Springer, 2020.

\bibitem{chatterjee2021visual}
Moitreya Chatterjee, Jonathan Le~Roux, Narendra Ahuja, and Anoop Cherian.
\newblock Visual scene graphs for audio source separation.
\newblock In {\em Proc. ICCV}, 2021.

\bibitem{cherian20222}
Anoop Cherian, Chiori Hori, Tim~K Marks, and Jonathan Le~Roux.
\newblock (2.5+ 1) d spatio-temporal scene graphs for video question answering.
\newblock In {\em Proceedings of the AAAI Conference on Artificial
  Intelligence}, volume~36, pages 444--453, 2022.

\bibitem{chung2014empirical}
Junyoung Chung, Caglar Gulcehre, KyungHyun Cho, and Yoshua Bengio.
\newblock Empirical evaluation of gated recurrent neural networks on sequence
  modeling.
\newblock {\em arXiv preprint arXiv:1412.3555}, 2014.

\bibitem{ephrat2018looking}
Ariel Ephrat, Inbar Mosseri, Oran Lang, Tali Dekel, Kevin Wilson, Avinatan
  Hassidim, William~T Freeman, and Michael Rubinstein.
\newblock Looking to listen at the cocktail party: a speaker-independent
  audio-visual model for speech separation.
\newblock {\em ACM Trans. Graph. (TOG)}, 37(4):1--11, 2018.

\bibitem{gan2020music}
Chuang Gan, Deng Huang, Hang Zhao, Joshua~B Tenenbaum, and Antonio Torralba.
\newblock Music gesture for visual sound separation.
\newblock In {\em Proc. CVPR}, pages 10478--10487, 2020.

\bibitem{gao20192}
Ruohan Gao and Kristen Grauman.
\newblock 2.5 d visual sound.
\newblock In {\em Proc. CVPR}, pages 324--333, 2019.

\bibitem{gao2019co}
Ruohan Gao and Kristen Grauman.
\newblock Co-separating sounds of visual objects.
\newblock In {\em Proc. ICCV}, pages 3879--3888, 2019.

\bibitem{geng2020spatio}
Shijie Geng, Peng Gao, Chiori Hori, Jonathan~Le Roux, and Anoop Cherian.
\newblock Spatio-temporal scene graphs for video dialog.
\newblock In {\em Proc. AAAI}, 2021.

\bibitem{hartley2003multiple}
Richard Hartley and Andrew Zisserman.
\newblock {\em Multiple view geometry in computer vision}.
\newblock Cambridge university press, 2003.

\bibitem{he2016deep}
Kaiming He, Xiangyu Zhang, Shaoqing Ren, and Jian Sun.
\newblock Deep residual learning for image recognition.
\newblock In {\em Proc. CVPR}, pages 770--778, 2016.

\bibitem{hershey1999audio}
John Hershey and Javier Movellan.
\newblock Audio vision: Using audio-visual synchrony to locate sounds.
\newblock In {\em Proc. NIPS}, pages 813--819, December 1999.

\bibitem{ioffe2015batch}
Sergey Ioffe and Christian Szegedy.
\newblock Batch normalization: Accelerating deep network training by reducing
  internal covariate shift.
\newblock In {\em International conference on machine learning}, pages
  448--456. PMLR, 2015.

\bibitem{jansson2017singing}
Andreas Jansson, Eric Humphrey, Nicola Montecchio, Rachel Bittner, Aparna
  Kumar, and Tillman Weyde.
\newblock Singing voice separation with deep {U}-net convolutional networks.
\newblock In {\em Proc. ISMIR}, October 2017.

\bibitem{ji2020action}
Jingwei Ji, Ranjay Krishna, Li~Fei-Fei, and Juan~Carlos Niebles.
\newblock Action genome: Actions as compositions of spatio-temporal scene
  graphs.
\newblock In {\em Proc. CVPR}, pages 10236--10247, 2020.

\bibitem{johnson2015image}
Justin Johnson, Ranjay Krishna, Michael Stark, Li-Jia Li, David Shamma, Michael
  Bernstein, and Li~Fei-Fei.
\newblock Image retrieval using scene graphs.
\newblock In {\em Proc. CVPR}, pages 3668--3678, 2015.

\bibitem{kidron2005pixels}
Einat Kidron, Yoav~Y Schechner, and Michael Elad.
\newblock Pixels that sound.
\newblock In {\em Proc. CVPR}, pages 88--95. IEEE, 2005.

\bibitem{kim2019audiocaps}
Chris~Dongjoo Kim, Byeongchang Kim, Hyunmin Lee, and Gunhee Kim.
\newblock Audiocaps: Generating captions for audios in the wild.
\newblock In {\em Proc. NAACL HLT}, pages 119--132, 2019.

\bibitem{kingma2014adam}
Diederik~P Kingma and Jimmy Ba.
\newblock Adam: A method for stochastic optimization.
\newblock In {\em Proc. ICLR}, 2014.

\bibitem{krasin2017openimages}
Ivan Krasin, Tom Duerig, Neil Alldrin, Vittorio Ferrari, Sami Abu-El-Haija,
  Alina Kuznetsova, Hassan Rom, Jasper Uijlings, Stefan Popov, Andreas Veit,
  et~al.
\newblock Openimages: A public dataset for large-scale multi-label and
  multi-class image classification.
\newblock {\em Dataset available from https://github. com/openimages}, 2(3):18,
  2017.

\bibitem{krishna2017visual}
Ranjay Krishna, Yuke Zhu, Oliver Groth, Justin Johnson, Kenji Hata, Joshua
  Kravitz, Stephanie Chen, Yannis Kalantidis, Li-Jia Li, David~A Shamma, et~al.
\newblock Visual {G}enome: Connecting language and vision using crowdsourced
  dense image annotations.
\newblock {\em Int. J. Comput. Vis.}, 123(1):32--73, 2017.

\bibitem{lee2019self}
Junhyun Lee, Inyeop Lee, and Jaewoo Kang.
\newblock Self-attention graph pooling.
\newblock In {\em Proc. ICML}, pages 3734--3743, June 2019.

\bibitem{li2021binaural}
Sijia Li, Shiguang Liu, and Dinesh Manocha.
\newblock Binaural audio generation via multi-task learning.
\newblock {\em ACM Transactions on Graphics (TOG)}, 40(6):1--13, 2021.

\bibitem{li2008modeling}
Xiaowei Li, Changchang Wu, Christopher Zach, Svetlana Lazebnik, and Jan-Michael
  Frahm.
\newblock Modeling and recognition of landmark image collections using iconic
  scene graphs.
\newblock In {\em Proc. ECCV}, pages 427--440. Springer, 2008.

\bibitem{liu2019divide}
Yuzhou Liu and DeLiang Wang.
\newblock Divide and conquer: A deep {CASA} approach to talker-independent
  monaural speaker separation.
\newblock {\em IEEE/ACM Trans. Audio, Speech, Lang. Process.},
  27(12):2092--2102, 2019.

\bibitem{lucas1981iterative}
Bruce~D Lucas, Takeo Kanade, et~al.
\newblock {\em An iterative image registration technique with an application to
  stereo vision}.
\newblock Vancouver, 1981.

\bibitem{luo2019fasnet}
Yi~Luo, Cong Han, Nima Mesgarani, Enea Ceolini, and Shih-Chii Liu.
\newblock Fasnet: Low-latency adaptive beamforming for multi-microphone audio
  processing.
\newblock In {\em 2019 IEEE automatic speech recognition and understanding
  workshop (ASRU)}, pages 260--267. IEEE, 2019.

\bibitem{meseguer2019conditioned}
Gabriel Meseguer-Brocal and Geoffroy Peeters.
\newblock Conditioned-{U}-{N}et: Introducing a control mechanism in the
  {U}-{N}et for multiple source separations.
\newblock {\em arXiv preprint arXiv:1907.01277}, 2019.

\bibitem{michelsanti2020overview}
Daniel Michelsanti, Zheng-Hua Tan, Shi-Xiong Zhang, Yong Xu, Meng Yu, Dong Yu,
  and Jesper Jensen.
\newblock An overview of deep-learning-based audio-visual speech enhancement
  and separation.
\newblock {\em arXiv preprint arXiv:2008.09586}, 2020.

\bibitem{morgado2018self}
Pedro Morgado, Nuno Vasconcelos, Timothy Langlois, and Oliver Wang.
\newblock Self-supervised generation of spatial audio for 360° video.
\newblock In {\em Proc. NeurIPS}, pages 360--370, 2018.

\bibitem{ochiai2020beam}
Tsubasa Ochiai, Marc Delcroix, Rintaro Ikeshita, Keisuke Kinoshita, Tomohiro
  Nakatani, and Shoko Araki.
\newblock Beam-tasnet: Time-domain audio separation network meets
  frequency-domain beamformer.
\newblock In {\em ICASSP 2020-2020 IEEE International Conference on Acoustics,
  Speech and Signal Processing (ICASSP)}, pages 6384--6388. IEEE, 2020.

\bibitem{owens2018audio}
Andrew Owens and Alexei~A Efros.
\newblock Audio-visual scene analysis with self-supervised multisensory
  features.
\newblock In {\em Proc. ECCV}, pages 631--648, 2018.

\bibitem{owens2016visually}
Andrew Owens, Phillip Isola, Josh McDermott, Antonio Torralba, Edward~H
  Adelson, and William~T Freeman.
\newblock Visually indicated sounds.
\newblock In {\em Proc. CVPR}, pages 2405--2413, 2016.

\bibitem{raffel2014mir_eval}
Colin Raffel, Brian McFee, Eric~J Humphrey, Justin Salamon, Oriol Nieto, Dawen
  Liang, Daniel~PW Ellis, and C~Colin Raffel.
\newblock mir\_eval: A transparent implementation of common mir metrics.
\newblock In {\em Proc. ISMIR}, 2014.

\bibitem{ranftl2021vision}
Ren{\'e} Ranftl, Alexey Bochkovskiy, and Vladlen Koltun.
\newblock Vision transformers for dense prediction.
\newblock In {\em Proc. ICCV}, 2021.

\bibitem{ren2016faster}
Shaoqing Ren, Kaiming He, Ross Girshick, and Jian Sun.
\newblock Faster r-cnn: towards real-time object detection with region proposal
  networks.
\newblock {\em IEEE Trans. Pattern Anal. Mach. Intell.}, 39(6):1137--1149,
  2016.

\bibitem{ronneberger2015u}
Olaf Ronneberger, Philipp Fischer, and Thomas Brox.
\newblock U-net: Convolutional networks for biomedical image segmentation.
\newblock In {\em Proc. MICCAI}, pages 234--241. Springer, 2015.

\bibitem{senocak2018learning}
Arda Senocak, Tae-Hyun Oh, Junsik Kim, Ming-Hsuan Yang, and In~So Kweon.
\newblock Learning to localize sound source in visual scenes.
\newblock In {\em Proc. CVPR}, pages 4358--4366, 2018.

\bibitem{slizovskaia2019end}
Olga Slizovskaia, Leo Kim, Gloria Haro, and Emilia Gomez.
\newblock End-to-end sound source separation conditioned on instrument labels.
\newblock In {\em Proc. ICASSP}, pages 306--310, May 2019.

\bibitem{subramanian2021directional}
Aswin~Shanmugam Subramanian, Chao Weng, Shinji Watanabe, Meng Yu, Yong Xu,
  Shi-Xiong Zhang, and Dong Yu.
\newblock Directional asr: A new paradigm for e2e multi-speaker speech
  recognition with source localization.
\newblock In {\em Proc. ICASSP}, pages 8433--8437. IEEE, 2021.

\bibitem{teed2020raft}
Zachary Teed and Jia Deng.
\newblock Raft: Recurrent all-pairs field transforms for optical flow.
\newblock In {\em Proc. ECCV}, pages 402--419. Springer, 2020.

\bibitem{tian2021cyclic}
Yapeng Tian, Di~Hu, and Chenliang Xu.
\newblock Cyclic co-learning of sounding object visual grounding and sound
  separation.
\newblock In {\em Proc. CVPR}, 2021.

\bibitem{tian2018audio}
Yapeng Tian, Jing Shi, Bochen Li, Zhiyao Duan, and Chenliang Xu.
\newblock Audio-visual event localization in unconstrained videos.
\newblock In {\em Proc. ECCV}, pages 247--263, 2018.

\bibitem{tzinis2020into}
Efthymios Tzinis, Scott Wisdom, Aren Jansen, Shawn Hershey, Tal Remez,
  Daniel~PW Ellis, and John~R Hershey.
\newblock Into the wild with {AudioScope}: Unsupervised audio-visual separation
  of on-screen sounds.
\newblock In {\em Proc. ICLR}, 2021.

\bibitem{velivckovic2017graph}
Petar Veli{\v{c}}kovi{\'c}, Guillem Cucurull, Arantxa Casanova, Adriana Romero,
  Pietro Lio, and Yoshua Bengio.
\newblock Graph attention networks.
\newblock In {\em Proc. ICLR}, April 2018.

\bibitem{Vincent2006BSSeval}
Emmanuel Vincent, R\'{e}mi Gribonval, and C\'{e}dric F\'{e}votte.
\newblock Performance measurement in blind audio source separation.
\newblock {\em IEEE Trans. Audio, Speech, Lang. Process.}, 14(4):1462--1469,
  July 2006.

\bibitem{wang2019dynamic}
Yue Wang, Yongbin Sun, Ziwei Liu, Sanjay~E Sarma, Michael~M Bronstein, and
  Justin~M Solomon.
\newblock Dynamic graph {CNN} for learning on point clouds.
\newblock {\em ACM Trans. Graph. (TOG)}, 38(5):1--12, 2019.

\bibitem{xu2021multiple}
Wenjie Xu, Maoshen Jia, Shang Gao, and Lu~Li.
\newblock Multiple sound source separation by using doa estimation and ica.
\newblock In {\em 2021 4th International Conference on Information
  Communication and Signal Processing (ICICSP)}, pages 249--253. IEEE, 2021.

\bibitem{xu2021visually}
Xudong Xu, Hang Zhou, Ziwei Liu, Bo~Dai, Xiaogang Wang, and Dahua Lin.
\newblock Visually informed binaural audio generation without binaural audios.
\newblock In {\em Proc. CVPR}, 2021.

\bibitem{zhao2020monocular}
Chaoqiang Zhao, Qiyu Sun, Chongzhen Zhang, Yang Tang, and Feng Qian.
\newblock Monocular depth estimation based on deep learning: An overview.
\newblock {\em Science China Technological Sciences}, 63(9):1612--1627, 2020.

\bibitem{zhao2019sound}
Hang Zhao, Chuang Gan, Wei-Chiu Ma, and Antonio Torralba.
\newblock The sound of motions.
\newblock In {\em Proc. ICCV}, pages 1735--1744, 2019.

\bibitem{zhao2018sound}
Hang Zhao, Chuang Gan, Andrew Rouditchenko, Carl Vondrick, Josh McDermott, and
  Antonio Torralba.
\newblock The sound of pixels.
\newblock In {\em Proc. ECCV}, pages 570--586, 2018.

\bibitem{zhou2018visual}
Yipin Zhou, Zhaowen Wang, Chen Fang, Trung Bui, and Tamara~L Berg.
\newblock Visual to sound: Generating natural sound for videos in the wild.
\newblock In {\em Proc. CVPR}, pages 3550--3558, 2018.

\end{thebibliography}
}
\appendix
\section{Details of the ASIW and AVE Datasets}
\label{datasets}
\noindent \textbf{Audio Separation in the Wild (ASIW) Dataset:} The \emph{Audio Separation in the Wild} (ASIW) dataset~\cite{chatterjee2021visual} consists of 147 validation, 322 test, and 10,540 training videos crawled from the larger AudioCaps dataset~\cite{kim2019audiocaps}. It has 14 auditory classes which are: \textit{baby, bell, bird, camera, clock, dogs, toilet/drain, horse, man/woman, telephone, train, sheep/goat, vehicle/bus, water/water tank}. Each video in the dataset is 10s long, has significant camera motion, and captures diverse audio-visual contexts. For the direction prediction task, we divide each video into temporal windows of 1s long consisting of 8 frames each and predict the direction of motion for a randomly chosen window.

\noindent \textbf{Audio Visual Event (AVE) Dataset}: The popular \emph{Audio Visual Event} (AVE) Dataset~\cite{tian2018audio} was originally designed for the task of identifying audio-visual events. We adapt it to evaluate our approach. We treat each audio-visual event class as an auditory class and associate with it a set of \emph{Auditory Objects}. The dataset contains 2211 training, 257 validation, and 261 test set videos. We use videos corresponding to 18 audio-visual event classes, including both potentially-moving and stationary object classes, for which pre-trained object detectors are avalable. The classes used for our experiments are: \textit{man, woman, car, plane, truck, motorcycle, train, clock, baby, bus, horse, toilet, violin, fiddle, flute, ukulele, acoustic guitar, banjo, accordion}. The videos in this dataset are 10s long as well. Similar to ASIW, we divide each video into windows of 1s long (i.e. 8 frames per window) for computing the motion displacement vectors and predict the direction of motion for a randomly chosen window.

\section{Generalizability of ASMP}
\label{generaizability}
To understand the real-world generalization of our approach to audio separation in  videos outside the datasets used for training the model, we downloaded several videos from Youtube each with an arbitrary mix of multiple sounds, with the goal of applying our approach on them to separate the sounds. We selected videos having audio in the set of classes of the ASIW dataset. We then applied the model trained on the ASIW dataset on these videos.  The results for these experiments are provided in the supplementary video. Our results clearly demonstrate that our approach leads to high quality audio separation even when the scene contains heavy distractors, such as continuous sound of rain, jammed traffic, or weak sounds, such as train announcements, etc. and exhibits reasonable amount of out of domain generalizability.

\section{Additional Ablation Results}
\label{sec:ablation}
In this section, we provide additional studies on the importance of each component in our model. We note that some of these results are already in the main paper, however are grouped in a different manner. 

\subsection{3D Graph and Direction Prediction}
In Table~\ref{tab:ablation_model}, we examine the contribution of each of the core novelties of our method on the ASIW dataset. \textbf{From the table, we see that the absence of the \emph{direction prediction loss} supervision deteriorates the SDR performance by 0.6 dB}. Given that SDR is in log-scale, this implies a significant drop in audio separation quality, thereby attesting to the importance of this loss in the separation task. In order to assess the contribution of the \emph{Pseudo 3D Scene Graph}, we replaced our 3D scene graph with a 2D scene graph (as in AVSGS~\cite{chatterjee2021visual}) 
, where the edges do not denote 3D spatial proximity, instead assume a fully-connected graph structure with equally-weighted edges. From the table, we see that this results in a drop of about $0.2$ dB in the SDR, emphasizing the criticality of this module. However, when the direction prediction loss is introduced back into the training regime, we notice a slight boost in performance. 

\begin{table}[t]
 \centering
 \sisetup{table-format=2.1,round-mode=places,round-precision=1,table-number-alignment = center,detect-weight=true} 
 \caption{SDR, SIR, SAR performance for different ablated model components of \name on the ASIW test set. [Best results in \textbf{bold}.] } 
 \resizebox{\linewidth}{!}{
 \begin{tabular}{rlSSS}
\toprule
\multirow{2}{*}{} & & \multicolumn{3}{c}{{ASIW}} \\ \cmidrule(l{0.25em}r{0.25em}){3-5}
{}& {Method}&  {SDR} $\uparrow$ & {SIR} $\uparrow$ &	{SAR} $\uparrow$  \\
\midrule
 1.~ & \name (Full Model) & \bfseries 9.6 &	\bfseries 14.5 & \bfseries	14.1   \\ \midrule
2.~ & \name: w/o Direction Prediction ($\lambda_4 = 0$) &  9.0 &	14.3 &	13.7    \\ 
3.~ & \name: w/o 3D Spatial Distances \& w/o Direction Prediction &  8.8 &	14.1 &	13.0    \\ 
4.~ & \name: w/o 3D Spatial Distances \& w/ Direction Prediction &  9.1 &	14.3 &	13.6    \\ 
\bottomrule
\end{tabular}
}
\label{tab:ablation_model}
\vspace*{0.3cm}
\end{table}

\subsection{Ablations on Losses}
Table~\ref{tab:ablation_loss} presents the results of ablating the different loss terms in our optimization objective on the ASIW dataset. The results reveal that the direction prediction loss alone, while crucial, is insufficient in singularly providing a reasonably accurate separation/direction prediction performance.  However, we find that training the models with any one of the other loss terms in conjunction with the direction prediction loss, results in a boost in model performance, with the presence of the co-separation loss resulting in the maximum gain.  This underscores the importance of all three loss terms: co-separation loss, consistency loss, and the orthogonality loss, besides the direction prediction loss for effective model training. 

\subsection{Compute Time Analysis}
Thanks to the use of GPUs, the Chamfer distance between pairs of point clouds can be computed very quickly. Typically, a forward pass of one batch (with 25 samples) through the full network takes about 1.2 seconds on a Intel Core i7 workstation with NVIDIA RTX 2070 GPUs. When the additional task of computing the Chamfer distance is not there, i.e. the 3D Scene Graph is replaced with a graph with equal edge weights, 
this results in a saving of only 0.06 seconds per batch.

\begin{table}[t]
 \centering
 \sisetup{table-format=2.1,round-mode=places,round-precision=1,table-number-alignment = center,detect-weight=true} 
 \caption{SDR, SIR, SAR, and Direction Prediction performance for loss ablated models on the ASIW test set. [Best results in \textbf{bold}.] } 
 \resizebox{\linewidth}{!}{
 \begin{tabular}{rlSSSS}
\toprule
\multirow{2}{*}{} & & \multicolumn{4}{c}{{ASIW}} \\ \cmidrule(l{0.25em}r{0.25em}){3-6}
{}& {Method}&  {SDR} $\uparrow$ & {SIR} $\uparrow$ &	{SAR} $\uparrow$  &	{10-class Dir. Acc. (in \%) } $\uparrow$ \\ 
\midrule
 1.~ & \name (Full Model) & \bfseries 9.55 &	\bfseries 14.5 & \bfseries	14.12 & \bfseries 42.51  \\ \midrule
2.~ & \name - Only Direction Prediction ($\lambda_1 = \lambda_2 = \lambda_3 = 0$) &  6.4 &	11.2 &	11.7 & 32.7  \\ 
3.~ & \name - Direction Prediction + $\mathcal{L}_{\mathrm{ortho}}$ ($\lambda_1 = \lambda_2 = 0$) &  6.42 &	11.65 &	10.1 & 33.2  \\ 
4.~ & \name - Direction Prediction + $\mathcal{L}_{\mathrm{co-sep}}$ ($\lambda_1 = \lambda_3 = 0$) &  7.91 &	13.24 &	10.88 & 35.6  \\ 
5.~ & \name - Direction Prediction + $\mathcal{L}_{\mathrm{cons}}$ ($\lambda_2 = \lambda_3 = 0$) &  6.39 &	11.61 &	11.93 & 33.1  \\ 
5.~ & \name - No Direction Prediction ($\lambda_4 = 0$) &  9.0 &	14.3 &	13.7 & \tdash \\
\bottomrule
\end{tabular}
}
\label{tab:ablation_loss}
\end{table}

\section{Performance Against Varying Kernel Scale}
\label{sec:rbf_ker}

 \begin{figure}[t]
    \centering
    \includegraphics[width=0.7\linewidth]{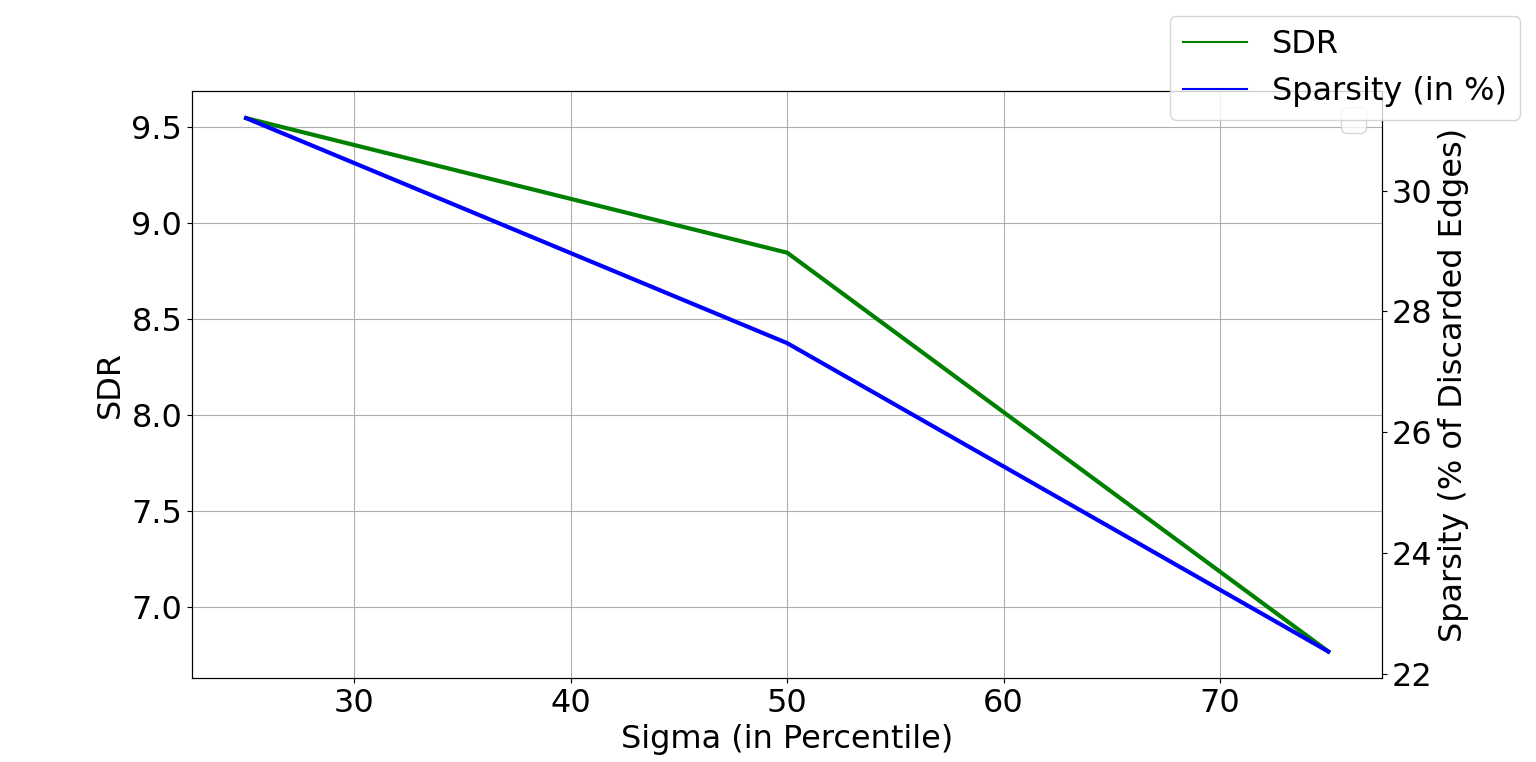}
   \caption{ A plot showing the sensitivity of our proposed \name model to the choice kernel bandwidth (percentile) against the sparsity induced into the graph.}
    \label{fig:sigma_sparse}
\end{figure}

Recall that the Radial Basis Function (RBF) kernel is used to model the weighted adjacency matrix of the 3D scene graph, where the kernel is constructed on the Chamfer distances between 3D coordinates of the graph nodes. Thus, the bandwidth of the kernel describes a \emph{soft ball} of a certain radius around each node to accommodate other nodes in the graph. As this radius depends on the specifics of the 3D graph constructed on the scene, it is not ideal to use a fixed radius for all graphs. Instead, we first compute a Chamfer distance matrix across all pairs of nodes in a given graph, and use the $\sigma$-th percentile of these distances to form the kernel bandwidth. In Figure~\ref{fig:sigma_sparse}, we plot the model performance (SDR) against increasing RBF kernel bandwidth, as well as the sparsity in the resulting graph (y-axis on the right), assessed by counting the number of edges with weights below $1e-5$ as a fraction of the total number of edges, which we discard. In the main paper, we reported results using the $25^{th}$-percentile of the Chamfer distance matrix, which corresponds to $\sigma=25\%$. In Figure~\ref{fig:sigma_sparse}, we plot against $\sigma=25, 50, 75$ as well. Note that using a higher $\sigma$ leads to several dense connections, resulting in lower sparsity, leading to mistaken contexts for a node for separation. We found that using $\sigma=25\%$ leads to the best performance.

\section{User Assessment Study}
\label{sec:user}

In order to subjectively assess the quality of audio source separation, we evaluated a randomly chosen subset of separated audio samples from \name and our closest competitor AVSGS~\cite{chatterjee2021visual} for human preferability on both ASIW and AVE datasets. For the purpose of this study, we chose a set of in-house annotators, who have successfully completed annotation/evaluation tasks for speech/audio separation, in the past. Table~\ref{tab:human_eval} reports these performances, which show a clear preference of the evaluators, for our method over AVSGS about 60--65\% of the time, on average. 

 \begin{table}[t]
 \centering
 \caption{Human preference score on samples obtained from our method vs. AVSGS~\cite{chatterjee2021visual}}\label{tab:human_eval}
 \begin{tabular}{lc}
 \toprule
 \textbf{Datasets} & \textbf{ Prefer ours} \\ 
 \midrule
 ASIW  - Ours vs. AVSGS~\cite{chatterjee2021visual}  & \textbf{63\%} \\ 
 AVE - Ours vs. AVSGS~\cite{chatterjee2021visual} & \textbf{68\%} \\ 
 \bottomrule
 \end{tabular}
 \end{table}

\section{Details of Compute Environment}
\label{sec:compute}

We conduct experiments on a cluster of workstations, each with Intel Core i7 CPU, with 256GB RAM. Each of the workstations is equipped with 8 NVIDIA RTX 2070 GPUs.  

\section{Network Architecture Details}
\label{sec:net_arch}

 \begin{figure}[t]
    \centering
    
    \includegraphics[width=\linewidth]{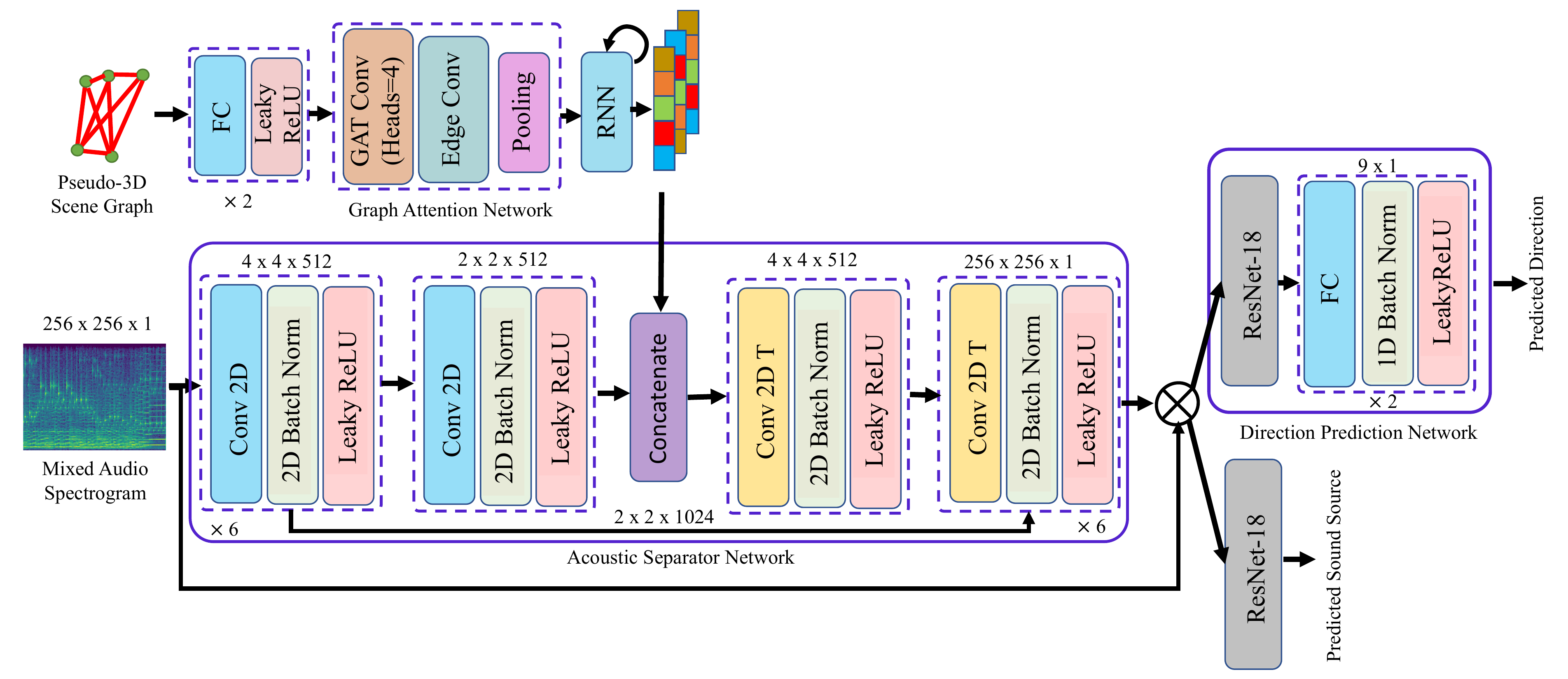} 
    
   \caption{A detailed illustration of our proposed \name model. }
    \label{fig:detail_arch}
\end{figure}

Our model, the \emph{\fullname}~(\name) has several components, as shown in Figure~\ref{fig:detail_arch}. Below, we list the key architectural details of each of them. 

\subsection{Feature Extractor}
Our model, \name, captures the visual representation of the objects in the scene by means of \textit{Pseudo-3D} scene graph; representing the visual entities of the scene as nodes of a graph (with associated features). These features are computed using a Faster R-CNN ($\frcnn$) network~\cite{ren2016faster}, with a ResNet-101~\cite{he2016deep} backbone pre-trained on the Visual Genome Dataset~\cite{krishna2017visual}. The features of the musical instrument type nodes that appear in the Audio Visual Event (AVE) dataset are obtained by training another $\frcnn$ network on the images of musical instruments on the OpenImages dataset~\cite{krasin2017openimages}. The $\frcnn$ network yields 2048-dimensional vectors for the Visual Genome dataset, which are then mapped to 512-dimensions via a 2-layer Multi-Layer Perceptron (MLP) with Leaky ReLU activations (negative slope=0.2), while the network yields 512-dimensional vectors for object features for the images in the OpenImages dataset and are used as is.

\subsection{Pseudo-3D Graph Attention Network}
Post the object detection and feature extraction, we construct the pseudo-3D scene-graph following the method laid out in the \textit{Proposed Method} section of the paper. The scene graph is processed by a \textit{Graph Attention Network}, which is a cascade of the following three modules:

\noindent \textbf{Graph Attention Network Convolution (GAT Conv)}: This module ~\cite{velivckovic2017graph} is responsible for updating the node features of the graph by employing a multi-headed graph message passing based on the adjacency of the nodes and edge weights. Our graphs are fully-connected and the edge weights are determined by the RBF kernel score as discussed in the paper. We design a 4-headed network, which outputs a 512-dimensional vector that embeds the full graph.

\noindent \textbf{Edge Convolution (Edge Conv)}: Edge Convolutions~\cite{wang2019dynamic} act on the output of the Graph Attention Network modules. These take in a concatenated pair of features associated with the two nodes of an edge and produces a vector as output. The input dimensions are thus $512 \times 2 = 1024$, while the output dimension is 512.

\noindent \textbf{Pooling Layers}: Finally, the \textit{Graph Attention Network} pools the updated features~\cite{lee2019self} across the nodes obtained from the previous steps. The Global Max and Average Pooling techniques are used for this purpose and the feature vectors from the two are concatenated.

\subsection{Recurrent Network}
To iteratively (and automatically) segment the graph into appropriate subgraphs, we introduce a Recurrent Network into the pipeline. We instantiate it using a \textit{Gated Recurrent Unit} (GRU)~\cite{chung2014empirical}, whose input space and feature dimensions are 512-dimensional.

\subsection{Acoustic Separator Network}
The actual task of separating a mixed audio into its constituents is undertaken by an encoder-decoder network called the \emph{Acoustic Separator Network}. In broad strokes, the network follows a U-Net~\cite{ronneberger2015u} style architecture, with an encoder, bottleneck, and decoder modules. The encoder and decoder parts consists of 7 convolution and  7 up-convolution layers, respectively. Each layer has $4 \times 4$ filters with LeakyRELU activations and negative slope of 0.2. Moreover, encoder layers with matching spatial resolution of output feature maps are connected by skip connections to the corresponding decoder layers. The bottleneck layer concatenates the encoder embedding with the scene graph embedding, with each element of the latter tiled $2 \times 2$ times. The dimensions of the bottleneck layer are $(2 \times 2 \times 512)$, each for the encoder embedding and the scene graph embedding.

\subsection{Direction Prediction Network}
A central innovation in \name is the aspect of using the motion direction of a sound source as an additional supervisory signal to train the audio separation module. In order to leverage this auxiliary supervision, we take time-slices of the separated source spectrogram and pass it through a network called the \emph{Direction Prediction Network}, which predicts which one of the 8 (or 26 depending on the setting) direction of motions (and one additional direction denoting no motion, and a second additional class denoting the displacement of the background) centered at the current location of the object, did the object move in, within the window. The network consists of a ResNet-18 style module~\cite{he2016deep} for embedding the separated spectrogram into a 512-dimensional vector. This is followed by a a 2-layer Multi-Layer Perceptron (MLP). The hidden layer embedding is passed through a 1d BatchNorm~\cite{ioffe2015batch} and Leaky ReLU activations (negative slope=0.2) before being processed by the second layer of the MLP. The output of the second layer is a 10 (or 28)-dimensional vector, constituting the logits of the direction prediction classifier.

\section{Qualitative Results}
\label{sec:qual_results}

\begin{figure}[t]
\centering
\includegraphics[width=\linewidth,trim={0cm 6.4cm 0cm 3cm},clip=True]{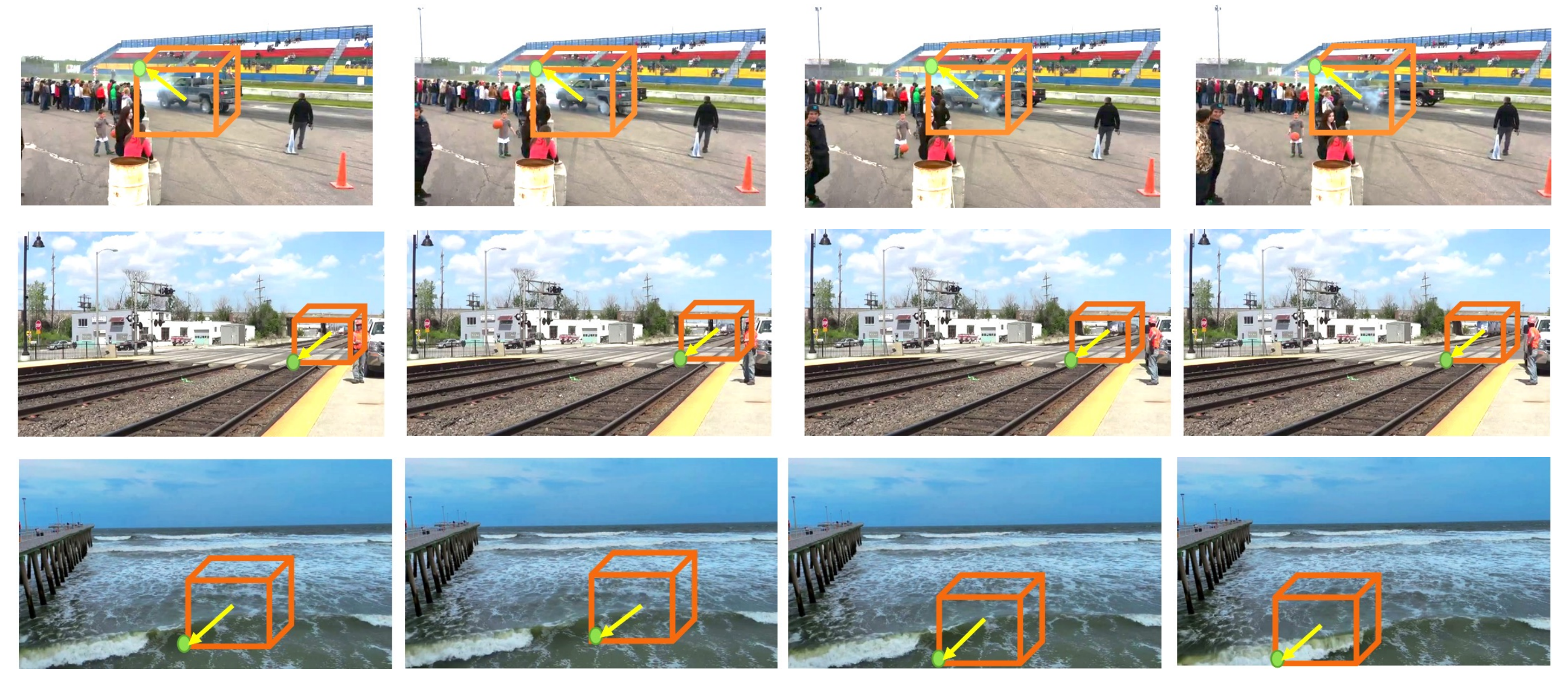}
  \includegraphics[width=\linewidth,trim={3cm 8cm 0cm 0cm},clip=True]{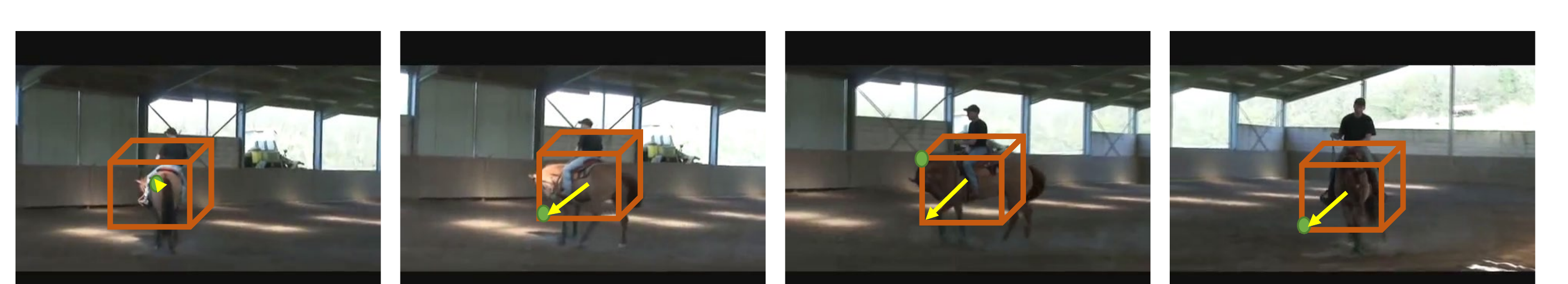}
\caption{Qualititive results from the ASIW dataset demonstrating object 3D motion prediction using our method. The \textcolor{green}{green} dot shows the ground truth direction on the \textcolor{orange} {orange} unit cube, and the \textcolor{yellow}{yellow} arrow shows the predicted direction. The unit cube is placed at the center of the object detection bounding box. 
} 
\label{fig:first_page}
\end{figure}

\begin{figure}[t]
    \centering
    \includegraphics[width=\linewidth,trim={3cm 0cm 0cm 8cm},clip=True]{figures/suppl_figs/ASIW/Sample_4.pdf}
    \includegraphics[width=\linewidth,trim={3cm 0cm 0cm 8cm},clip=True]{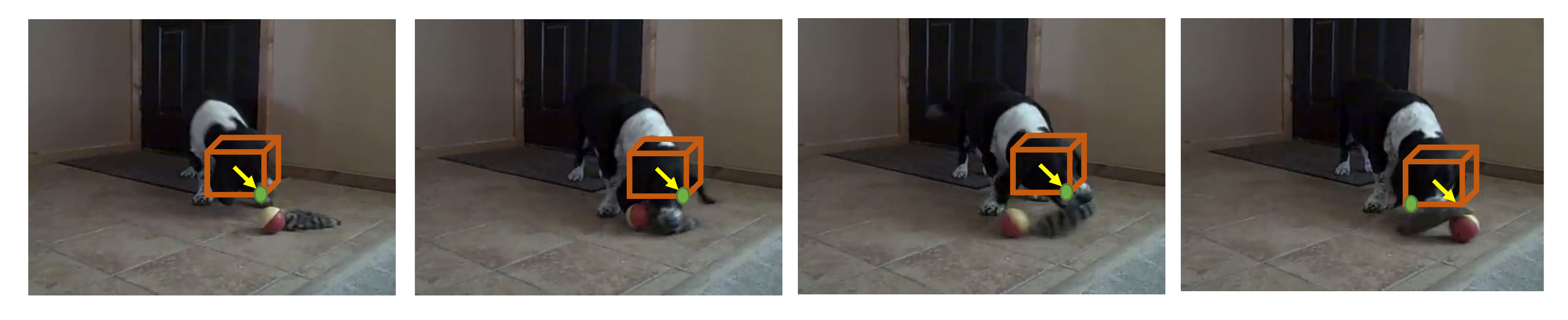}
    \includegraphics[width=\linewidth,trim={3.5cm 1cm 2.5cm 8cm},clip=True]{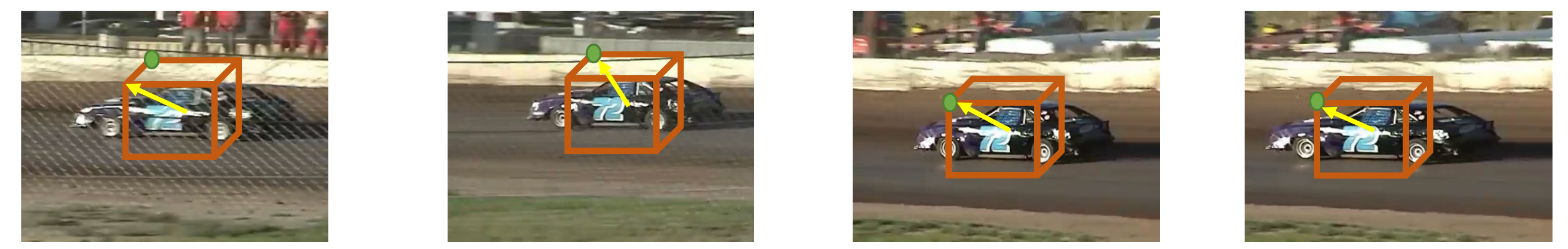}
    \includegraphics[width=\linewidth,trim={3cm 8cm 0cm 0cm},clip=True]{figures/suppl_figs/ASIW/Sample_7.pdf}
    \includegraphics[width=\linewidth,trim={3cm 8cm 0cm 0cm},clip=True]{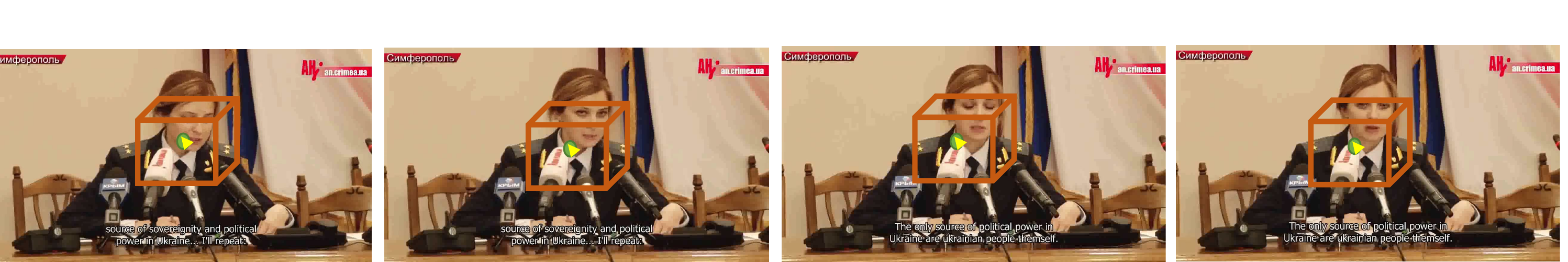}
    \includegraphics[width=\linewidth,trim={3cm 0cm 0cm 8cm},clip=True]{figures/suppl_figs/AVE/Sample_2.pdf}
   \caption{Additional qualitative direction prediction results on ASIW videos. We show a unit cube around the \emph{Auditory Object} with a green dot denoting the ground truth direction of motion. The yellow arrow indicates the predicted motion direction by \name using only the audio signal from a mixed spectrogram.}
    \label{fig:ave_quals}
\end{figure}

\begin{figure}[t]
    \centering
    \includegraphics[width=\linewidth,trim={3cm 8cm 0cm 0cm},clip=True]{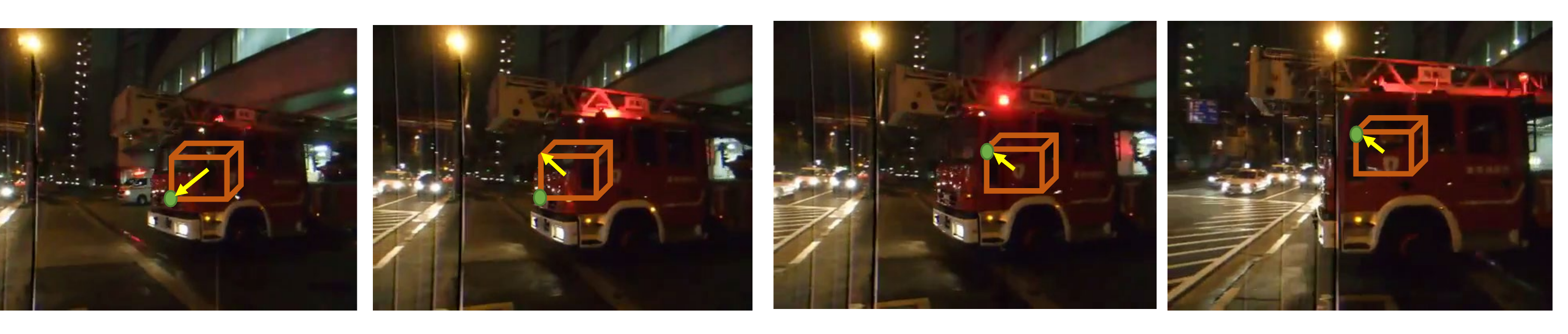} 
    \includegraphics[width=\linewidth,trim={3cm 0cm 0cm 8cm},clip=True]{figures/suppl_figs/AVE/Sample_5.pdf} 

    \includegraphics[width=\linewidth,trim={3cm 0cm 0cm 8cm},clip=True]{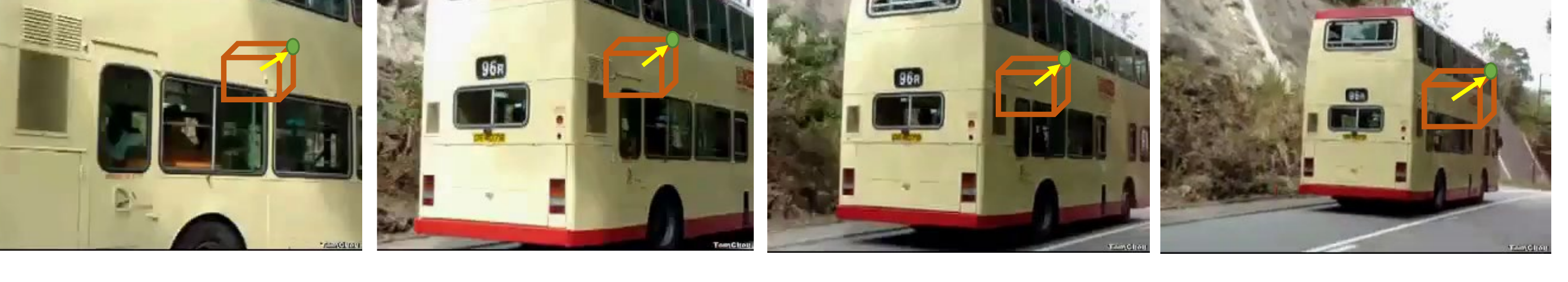}
    
     \includegraphics[width=\linewidth,trim={3cm 8cm 0cm 0cm},clip=True]{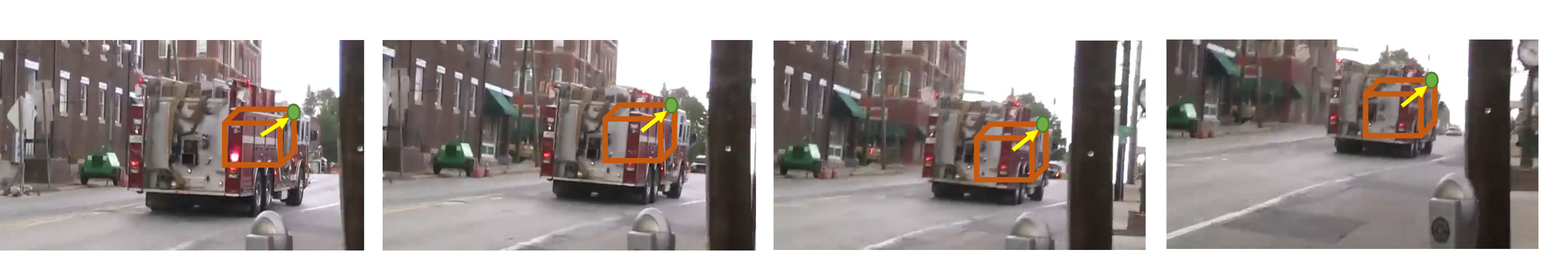}
    \includegraphics[width=\linewidth,trim={3cm 0cm 0cm 8cm},clip=True]{figures/suppl_figs/AVE/Sample_4.pdf}
    
    \includegraphics[width=\linewidth,trim={3cm 8cm 0cm 0cm},clip=True]{figures/suppl_figs/AVE/Sample_5.pdf}
    \includegraphics[width=\linewidth,trim={3cm 0cm 0cm 8cm},clip=True]{figures/suppl_figs/AVE/Sample_5.pdf}
   \caption{Qualitative direction prediction results on AVE videos. We show a unit cube around the \emph{Auditory Object} with a green dot denoting the ground truth direction of motion. The yellow arrow indicates the predicted motion direction by \name using only the audio signal from a mixed spectrogram.}
    \label{fig:asiw_quals}
\end{figure}

In Figures~\ref{fig:first_page},~\ref{fig:asiw_quals}, and~\ref{fig:ave_quals}, we present several qualitative results demonstrating the direction prediction by \name across ASIW and AVE datasets. We show these results on the starting frame of a window used for the direction prediction, and show a sequence of such windows.  Also shown is a cube centered at an \emph{Auditory Object} in the scene, with one of its corners marked with a green dot denoting the true direction of motion of the object in the window (as estimated using our optical flow estimation, between the first and the last frames in a window, and combined with the pseudo 3D depth estimation. Note that, if an object has no motion, the green dot is then placed at the center of the cube. The estimated motion directions are denoted by yellow-colored vectors and point to one of the eight corners of the cube or to the cube center (in case there is no motion). 

\end{document}